\documentclass[12pt]{article}
\usepackage{epsfig}

\parskip=\medskipamount	
\renewcommand{\thesection}{\arabic{section}}

\catcode`@=11

\@addtoreset{equation}{section}
\@addtoreset{equation}{subsection}
\def\theequation{\ifnum\value{section}=0 \arabic{equation}\ignorespaces
\else \ifnum\value{section}=-1 A.\arabic{equation}\ignorespaces
\else \ifnum\value{subsection}=0 \thesection.\arabic{equation}\ignorespaces
\else \thesection.\arabic{subsection}.\arabic{equation}\ignorespaces
                             \fi
                        \fi
                   \fi}

{\catcode`\'=\active \def'{{}^\bgroup\prim@s}}

\catcode`@=12

\newcommand{\bq}{\begin{equation}}
\newcommand{\be}{\begin{equation}} 
\newcommand{\fq}{\end{equation}}
\newcommand{\ee}{\end{equation}}
\newcommand{\bqr}{\begin{eqnarray}}
\newcommand{\beqs}{\begin{eqnarray}} 
\newcommand{\fqr}{\end{eqnarray}}
\newcommand{\eeqs}{\end{eqnarray}}
\newcommand{\non}{\nonumber \\[1ex]}

\newcommand{\rf}[1]{(\ref{#1})}  

\newcommand{\una}{{\underline a}} 
\newcommand{\unb}{{\underline b}}

 \def\cN{{\cal N}}

\def\pa{\partial}

\def\half{{1\over 2}}

\def\on#1#2{{\buildrel{\mkern2.5mu#1\mkern-2.5mu}\over{#2}}}
\def\dt#1{\on{\hbox{\bf .}}{#1}}                

\def\a{\alpha} 
\def\b{\beta} 
\def\da{\dt\alpha}
\def\db{\dt\beta}

\def\bop#1{\setbox0=\hbox{$#1M$}\mkern1.5mu
	\vbox{\hrule height0pt depth.04\ht0
	\hbox{\vrule width.04\ht0 height.9\ht0 \kern.9\ht0
	\vrule width.04\ht0}\hrule height.04\ht0}\mkern1.5mu}
\def\Box{{\mathpalette\bop{}}}                        

\begin{document} 

\thispagestyle{empty}

\begin{flushright}
\begin{tabular}{l}
ANL-HEP-PR-00-117 \\ 
YITP-SB-00-72\\ 
hep-th/0101025 \\ 
\end{tabular}
\end{flushright}

\vskip.3in
\begin{center}
{\Large\bf Simplifying Algebra in Feynman Graphs
\vskip .2cm
Part III: Massive Vectors}
\vskip.3in 

{\bf Gordon Chalmers}
\\[5mm]
{\em Argonne National Laboratory \\
High Energy Physics Division \\
9700 South Cass Avenue \\
Argonne, IL 60439-4815 } \\
{Email: chalmers@pcl9.hep.anl.gov}\\[5mm]

\vskip .2in
{\bf Warren Siegel}
\\[5mm]
{\em C.N. Yang Institute for Theoretical Physics \\
State University of New York at Stony Brook \\
Stony Brook, NY 11794-3840}\\
{Email: siegel@insti.physics.sunysb.edu}

\vskip.5in minus.2in

{\bf Abstract}

\end{center}

A T-dualized selfdual inspired formulation of massive vector fields coupled to 
arbitrary matter is generated; subsequently its perturbative series modeling a 
spontaneously broken gauge theory is analyzed.  The new Feynman rules and 
external line factors are chirally minimized in the sense that only one type 
of spin index occurs in the rules.  Several processes are examined in detail 
and the 
cross-sections formulated in this approach.  A double line formulation of the 
Lorentz algebra for Feynman diagrams is produced in this formalism, similar to 
color ordering, which follows from a spin ordering of the Feynman rules.  The 
new double line formalism leads to further minimization of gauge invariant 
scattering in perturbation theory.  The dualized electroweak model is also 
generated. 
 
\setcounter{page}{0}  
\newpage 
\setcounter{footnote}{0}


\section{Introduction}

Much progress has been made in the development of new tools useful for
performing perturbative gauge theory calculations at high orders.  These techniques 
include string inspired ones \cite{Bern:1991cu,Bern:1992aq}, color ordering
\cite{Bern:1991ux,Bern:1994zx}, spinor helicity \cite{spinorhel,spinorhell,oldtwist}, 
constraints based on analyticity \cite{Chalmers:1994th, 
Bern:1995ix}, unitarity methods \cite{Bern:1994zx}, and those based on 
selfdual field theories \cite{Chalmers:1996rq,Chalmers:1999ui,Chalmers:1999jb, 
Chalmers:1997sg}.  These methods have enabled the calculation of several 
higher-point amplitudes necessary for next-to-leading order phenomenology as 
well as several sequences of helicity amplitudes with an arbitrary number of 
external legs and in multiple dimensions.\footnote{See \cite{Mangano:1991by} 
and \cite{Bern:1996je} for reviews at tree and loop level.  Exact sequences 
of amplitudes include those in \cite{Bern:1994zx,ptamps,Bern:1994qk,Mahlon:1994si, 
Bern:1995cg,Bern:1998xc,Bern:1997ja}.}

In \cite{Morgan:1995te} a second-order formulation of fermionic couplings is 
derived by giving a chiral reduction of the minimally coupled theory.  Integrating 
out the ${\bar\psi}^{\da}$ and ${\bar\chi}^{\da}$ components in the 
minimally coupled action $S={\rm Tr} \int d^4x {\cal L}$
(in the conventions of {\it Superspace} \cite{Superspaceyoufool}), 
\be 
{\cal L}= -{\bar\psi}^{\da} i\nabla_{\a\da} \psi^\a 
 -{\bar\chi}^{\da} i\nabla_{\a\da} \chi^{\a} +
m(\psi^\a\chi_\a  + {\bar\psi}^{\da}{\bar\chi}_{\da} )
\ee
leads to a Lagrangian 
\be 
{\tilde{\cal L}}= -\psi^\a (\Box - m^2) \chi_\a + \psi^\a
F_\a{}^\b \chi_\b \ ,
\label{reduce}
\ee
and a gauge covariantized theory where: (1) only the
selfdual component of the  gauge field strength couples to the
fermions, and (2) in the remaining couplings the fermions are 
effectively bosonized. The new Lagrangian leads to a reduction in the 
amount of algebra one normally encounters in the computation of amplitudes
because, for one reason, of the elimination of the gamma matrices
directly at the level of the Lagrangian.  Furthermore, because of the
coupling only to the selfdual field strength \cite{Chalmers:1996rq}, 
in MHV amplitudes only the first terms in \rf{reduce} contribute and a 
perturbation around MHV structure is obtained through insertions of the 
latter terms.   This is seen from the Lagrangian in eq.~(\ref{reduce}) 
from the spin independence when truncating to $F_{\a\b}=0$ (i.e.\ for 
amplitudes between states of the same helicity) \cite{Chalmers:1999ui}.  
Because of the spin independence of the couplings in this case, known 
supersymmetry identities \cite{supid,Mangano:1991by} that these amplitudes obey 
become trivial.  These properties are also a consequence of spectral 
flow in the $\cN=2$ string theories \cite{Chalmers:2000bh}.  

In \cite{Chalmers:1999jb} we expanded and generalized these tools to the case
of all partons, with spin $\leq 1$, by introducing a
``spacecone'' analogous to the well known lightcone, but with the
spinor helicity technique incorporated through the use of external line 
factors in a complex gauge.   Amplitudes are defined with
the choice to identify the two lightlike axes with two of the physical
external momenta in accord with the reference momenta choice in the
spinor helicity method.  The progress along these lines arises by
manipulating directly the theory in the Lagrangian and
treating the gauge field lines as scalar components.  
Amplitudes in such theories
closer to maximally helicity violating (for spin $1$) are also easier
to calculate than in  previous formulations:  For example,
in the massless theory even at five-point the helicity
structure is such that all amplitudes are within two helicity flips
to the maximal case.  
In these two
applications, the known supersymmetry identities  become extremely simple,
and the closer one is to selfdual helicity configurations  the
simpler the amplitude calculations are as the couplings in the MHV 
limit are scalar-like.

Selfdual Yang-Mills field theories may be formulated in a Lorentz 
covariant fashion through 
\bqr  
{\cal L}= {\rm Tr}~ G^{\a\b} F_{\a\b} \ , \qquad  F_{\a\b} = 
{i\over 2} \partial_{(\a}{}^{\dt\gamma} A_{\b)\dt\gamma} + A_\a{}^{\dt\gamma} 
A_{\b\dt\gamma} \ , 
\label{sdlorentz}
\fqr 
where $G_{\a\b}$ is a multiplier that enforces the selfdual field equations.  
Different gauge-fixed versions of this action as well as one-field 
non-Lorentz covariant actions are found in the references 
(e.g.\ \cite{Chalmers:1996rq}).   
Gauge theories may be formulated as a perturbation around this Lorentz 
covariant action through the addition of the term $G^{\a\b} G_{\a\b}$ 
followed by integrating over the multipliers.  ${\cal N}=2$ string theory also 
admits a description of selfdual and quantum supersymmetric selfdual 
(the latter possessing a trivial S-matrix) description and possesses a 
gauge theory correspondence with the MHV\footnote{MHV represents maximally 
helicity violating amplitudes for the reason that all partons of out-going 
momentum have the same helicity.} sector.  The perturbative reformulation of 
gauge theory coupled to matter of various spins near the selfdual point admits 
improved diagrammatics for calculating amplitudes; in addition, certain 
chirality properties of the reformulated gauge theory are made manifest.  In 
this work we examine the formulation of massive theories via dualization of 
perturbed selfdual theories.  

The connection between selfduality and quantum field theory in both 
the classical and quantum regime allows a mapping between the dualized 
formulations to the $\cN=2$ string theories, in both the bosonic and 
target spacetime supersymmetric versions.  The anomalous quantum one-loop 
amplitudes obtained via sewing the finite number of particles in the 
the $\cN=2$ string agree with those of the MHV amplitudes \cite{Chalmers:2000bh}.  
(The amplitudes are anomalous because they lack a modular invariant integration 
measure.  Otherwise the RNS amplitudes would equal zero at 
genus one, in either ${\cal N}=2$ \cite{Chalmers:2000wd} or $\cN=4$ form 
\cite{Berkovits:1995vy}.)  The supersymmetric extension 
analyzed in \cite{Chalmers:2000bh} via superselection sectors generates a 
trivial S-matrix, matching with those of supersymmetric selfdual 
theories to all loop orders in the covariant version \cite{Chalmers:1996rq}.  
Perturbation in helicity about the selfdual point allows a direct connection 
between the $\cN=2$ string amplitudes and the selfdual sector of gauge and 
gravity theory \cite{Chalmers:1996rq}.   

Many of the ideas behind these works arise in formulating Yang-Mills theory 
as a perturbation around the selfdual limit, in the sense that the former 
is described through the first order formulation
\be
{\cal L} = {\rm Tr}~\Bigl(-{g^2\over 2} G^{\a\b}G_{\a\b} +
G^{\a\b}F_{\a\b} + {1\over 2} M^2 A^{\a\da} A_{\a\da} 
\Bigr)  \ , 
\label{dualmth}
\ee
which upon integrating the $A^{\a\da}$ field generates a simultaneously 
unitarity and renormalizable gauge theory. (The mass term in \rf{dualmth} 
is obtained via spontaneous symmetry breaking, described in this work.)  
In the massless limit the lowest order in $g$ term ($GF$) describes a 
two-field selfdual Yang-Mills model proposed in \cite{Siegel:1992xp}.  
It was perturbatively quantized and solved in \cite{Chalmers:1996rq}, 
generating the color-ordered MHV one-loop S-matrix, 
\bqr 
A_{n;1}(k_1^+,\ldots,k_n^+) = -{i\over 48\pi^2} 
 \sum_{1\leq i<j<k<l\leq n} {\langle ij\rangle [jk] \langle kl\rangle 
 [li] \over \langle 12\rangle \langle 23\rangle \cdots \langle n1\rangle} 
 \ ,
\label{MHVoneloop} 
\fqr 
together with a non-vanishing three-point vertex in $d=2+2$ dimensions.  
Here we listed the leading-in-color contribution for an adjoint 
vector in the loop with color structure (analogous to attaching Chan-Paton 
factors to the boundaries of the annulus), 
\bqr  
{\rm Tr}~ T^{\sigma_1} \ldots T^{\sigma_n} \ , 
\fqr     
where the individual $T^{\sigma_j}$ are the color matrices for the $j$th 
external line. (See \cite{Bardeen:1996gk,Cangemi:1997rx} for related work 
and \cite{sdfield} 
for formulations based on non-Lorentz covariant actions.)  After integrating
out the gauge fields $A^{\a\db}$ rather than the tensors
$G^{\a\b}$ one arrives at a dual formulation of a gauge theory, written 
as a sigma  model in the tensors $G^{\a\b}$ \cite{Chalmers:1997sg}, and in 
the scalars that generated the mass term.  
(Related work includes the dualized models \cite{Freedman:1981us}.)
The latter theory also may be used to efficiently derive the amplitudes closer 
to maximally helicity violating (or selfdual); it furthermore has several 
applications, from its formulation as a non-linear sigma model of Yang-Mills 
theory.  
In this work we explore its application as a dual formulation of
spontaneously broken gauge theories.  

The outline of this work is the following.  In Section 2 we present several 
examples within the dualized abelian theory (i.e.\  Stueckelberg model).  The 
dual theory of the Fermi interactions are obtained naturally in our formulation.  
In Section 3 we examine general reduction associated with non-abelian models 
and derive the corresponding dual formulations.  In Section 4 we examine 
the new Feynman rules of the dualized non-abelian theories.  In Section 5 
derive the analog of color flow but for the Lorentz group, i.e.\ spin ordering.  
We give a four-point massive vector amplitude in Secton 6 as an example.  In 
Section 7 we give the spin ordering in vector notation.  Section 8 contains the 
derivation of cross-sections in the dual formulation.  In section 9 we examine 
the (Glashow-Salam-Weinberg) electroweak model.  In Section 10 we end with a 
discussion of further 
relevant work associated with the selfdual and dualized massive vector theories 
described here. 

\section{Massive QED} 

\subsection{Feynman rules}

We first briefly review the dualization of an abelian theory 
and give an example scattering process.  We also describe the line 
factors associated with the massive states incorporating spinor 
helicity techniques.  

In \cite{Chalmers:1997sg} we considered the dualization of abelian vector 
fields.  We begin with a Stueckelberg theory, 
\be  
{\cal L} = {1\over 2} F^{\a\b}F_{\a\b} + {M^2\over 2}  
A^{\a\da} A_{\a\da} \ , 
\label{ymlag}
\ee 
where $F^{\a\b}$ is the selfdual field strength, 
\be  
F_{\a\b} = {i\over 2} \partial_{(\a}{}^{\db} A_{\b)\db} 
\ ,
\ee 
We also include the fermions through the minimal coupling 
\be  
{\cal L}_\psi= -{\bar\psi}^{\da} i\nabla_{\a\da} \psi^\a 
- {\bar\chi}^{\da} i\nabla_{\a\da} \chi^{\a} +
m(\psi^\a\chi_\a  + {\bar\psi}^{\da}{\bar\chi}_{\da} ) \ .
\label{minimalcoup}
\ee   
For completeness, we will also discuss the free, neutral analog,
\be  
{\cal L}_\psi= -{1\over 2}{\bar\psi}^{\da} i\partial_{\a\da} \psi^\a 
+{1\over 2}
m(\psi^\a\psi_\a  + {\bar\psi}^{\da}{\bar\psi}_{\da} ) 
\ee   
for purposes of comparing external line factors.  (Of course, in the 
nonabelian case this 
action can be coupled in real representations.)  We may express the Lagrangian in 
(\ref{ymlag}) in first order form with the addition of the fields $G^{\a\b}$, 
\be  
{\cal L} = -{1\over 2} G^{\a\b} G_{\a\b} 
+ G^{\a\b}F_{\a\b} + {M^2\over 2}  
A^{\a\da} A_{\a\da} \ . 
\label{firstord} 
\ee 
Now the action is quadratic in the original gauge field 
$A^{\a\da}$, and we may eliminate $A$ in exchange for $G$
(after integrating by parts the 
$\partial A$ term in the selfdual field strength).  Furthermore, we may 
also integrate out all of the barred components of the fermions as 
the action is also quadratic in these fields.  

In doing so, we obtain the dualized theory, 
\bqr   
{\cal L}_{\rm red} &=& -{M^2 \over 2} G^{\a\b} G_{\a\b} 
- \psi^\a \left( \Box - m^2\right) \chi_\a 
\non & &  
- {1\over 2} {1\over M^2-\psi^\a \chi_\a}  
\Bigl( \psi^\b \partial_{\a\da} \chi_\b + {1\over 2} 
\partial^\b{}_{\da} \left[ \psi_{(\a} \chi_{\b)}        
+ 2M G_{\a\b}\right]\Bigr)  
\non & & \times 
\Bigl( \psi^\rho \partial^{\a\da} \chi_\rho + {1\over 2} 
\partial^{\rho\da} \left[ \psi^{(\a} \chi_{\rho)}  
+ 2M G^\a{}_\rho \right] \Bigr) \ .
\label{red} 
\fqr   
We shall drop the Jacobians from the integration, which
vanish in dimensional reduction (regularization).  We further rescaled the fields  
$\psi\rightarrow \sqrt{2m} \psi$, $\chi\rightarrow\sqrt{2m} \chi$ and 
$G_{\a\b} \rightarrow M G_{\a\b}$ to simplify the 
coefficients.\footnote{We denote the mass of fermions with lower case 
$m$ and the mass of vectors with upper case $M$ in this work.}  Similar 
reductions may be performed in the Abelian Higgs model.  The reduced 
theory in (\ref{red}) for the massive particles is simplified in that 
there is no gamma matrix algebra and only one type of spin index occurs 
labeling the particle content for both the fermions and gauge bosons.  
The propagators are of second order form and the dimensions of the 
{\it rescaled} fields are $[G]=[\psi]=[\chi]=1$.  

The Feynman rules from the dualized theory in (\ref{red}) 
contain the propagator for the vectors $G^{\a\b}$ and 
for the fermion and are both second order.  Expanding the 
inverse of $1-\left({e\over m}\right)^2 \psi^\a\chi_\a$ produces 
an infinite number of terms.  We list the two vertices contributing 
to the $\psi^2 G^2$ process which entails expanding the action in 
(\ref{red}) to second order in the fermions: 
\bqr  
{\cal L}_{\rm red} &=& {1\over 2} G^{\a\b}\left( \Box - M^2 
\right) G_{\a\b} - \psi^\a \left( \Box - m^2\right) 
\chi_\a 
\non & &  
-{1\over M} \psi^\b \partial_{\a\da} \chi_\b 
\partial^{\rho\da} G^\a{}_\rho + {1\over 2M} G^{\a\b} 
\Box \left( \psi_{(\a} \chi_{\b)} \right) 
\non & &   
+{1\over 2 M^2} \psi^\a \chi_\a \left( \partial^\b{}_{\da} 
G_{\a\b}\right) \left(\partial^{\rho\da} G^\a{}_\rho\right) 
-{1\over 2M^2} \left( \psi^\b\partial_{\a\da} \chi_\b\right) 
 \left(\psi^\rho \partial^{\a\da}\chi_\rho\right)  
\non & &  
-{1\over 2M^2} \psi^\b\partial_{\a\da}\chi_\b 
\partial^{\rho\da} \left[ \psi^{(\a}\chi_{\rho)} \right] 
+{1\over 8M^2} \psi^{(\a} \chi^{\b)} \Box \left( \psi_{(\a} 
\chi_{\b)}\right) \ . 
\label{expanded} 
\fqr  
The propagator for $G^{\a\b}$ is given by 
\bqr  
\Delta^{\a\b,\gamma\rho}(k) = {1\over k^2+M^2}~ \Bigl[ 
 C^{\a\gamma} C^{\b\rho} + C^{\a\rho} C^{\b\gamma} \Bigr] \ . 
\fqr 
The three-point 
vertex found from the expansion in \rf{expanded} is,   
\be  
\langle \psi_\a (k_1) \chi_\b (k_2) G_{\mu\nu}(k_3) \rangle = 
 -{1\over M} \left( C_{\a\b} (k_2 k_3)_{\mu\nu} + {1\over 2} k_3^2 
C_{\b (\mu} C_{\nu)\a} \right) \ ,
\label{ruleone} 
\ee 
in which the second term is also symmetric in $\a,\b$, 
together with the $\psi^2 G^2$ one, 
\be 
\langle \psi_\a (k_1) \chi_\b (k_2) G_{\mu_1\nu_1}(k_3) 
G^{\mu_2\nu_2}(k_4) \rangle = - {1\over 4 M^2} 
C_{\a\b} (k_3 k_4)_{(\mu_1}{}^{(\mu_2} C_{\nu_1)}{}^{\nu_2)} \ ,
\label{ruletwo} 
\ee 
where we have defined the shorthand notation 
\be  
(kp)^{\a\b} = k^{\a\da} 
p^\b{}_\da \ . 
\ee 
Note that only the {\it undotted} indices appear explicitly in the 
Feynman rules.  
The four-point vertex corresponding to $\psi^2\chi^2$ is in k-space, 
$$ 
\langle \chi_{\a_1}(k_1) \psi_{\b_1}(k_2) \chi_{\a_2}(k_3) 
 \psi_{\b_2}(k_4) \rangle = 
$$
\be
-{1\over 2 M^2} \left[ 
 k_1\cdot k_3 ~C_{\a_1[\b_1} C_{\b_2]\a_2} 
+ (k_1,k_3+k_4)_{\a_2[\b_2} C_{\b_1]\a_1} 
- {1\over 4} (k_1+k_2)^2 C_{\b_1(\b_2} C_{\a_2)\a_1} \right] \ . 
\label{fourpointvertex}
\ee 
Unlike the undualized theory, there are explicit four-fermion vertices 
appearing in the rewriting of the original massive QED.

\subsection{External line factors}

We next specify the line factors for the massive fermions and vectors 
through the use of spinor helicity techniques \cite{Kife}.
Our conventions are as follows: In the {\it massless} case, Weyl spinors 
of momentum $k$ in $d=3+1$ are labeled 
according to $\psi^\a(k)=k^\a$ and the conjugate ${\bar\chi}^\da(k)=k^\da$.  
(In $d=2+2$ they are independent.) 
Each massless momentum $k^{\a\da}$ is associated with a twistor as
$k^{\a\da}=k^\a k^\da$ for positive energy, or $-k^\a k^\da$ for negative.
These invariants satisfy 
$s_{ij}=(k_i+k_j)^2=-\langle ij\rangle [ji]$
in terms of the spinor products $\langle ij\rangle$ of the undotted twistors 
and $[ij]$ of the dotted.  

The {\it massive} theory in \rf{minimalcoup} generates the field equation 
$k_{\a\da} \psi^\a + m\chi_\da=0$, 
and its complex conjugate, that possesses two solutions at momentum 
$k$ for the fermion.  In the complex case there are two independent 
solutions at 
momentum $-k$, while in the real case ($\chi=\psi$) they are related.  
The solutions 
may be specified in terms of {\it two} Weyl spinors of momentum $k_{(+)}$ 
and $k_{(-)}$ such that the momentum 
\be
k^{\a\da} = k_{(+)}^{\a\da} + k_{(-)}^{\a\da} 
= k_{(+)}^\a k_{(+)}^\da + k_{(-)}^\a k_{(-)}^\da \ , 
\qquad m^2 = -k^2 = \langle +-\rangle [- +] 
\label{spinorline}
\ee  
for positive energy (and $k=-k_{(+)}-k_{(-)}$ for negative energy). 
The ``spin vector" $S^{\a\da}$, satisfying
\be
S^2 = 1, \quad k\cdot S = 0
\ee
 whose spatial part defines the axis with respect to which 
$k_{(\pm)}^\b$ describe states of $s_z=\pm\half$, is then
\be
mS^{\a\da} = k_{(+)}^{\a\da} - k_{(-)}^{\a\da}
\ee

Since there are two solutions for the two-component spinor, 
the choice of basis is arbitrary:  
The simplest choice is obviously to choose a basis
proportional to $k_{(\pm)}^\b$, 
\be
\psi_{(\pm)}^\b = \mu k_{(\pm)}^\b
\qquad\Rightarrow\qquad 
\psi_{(+)}^\a \psi_{(-)}^\b - \psi_{(-)}^\a \psi_{(+)}^\b =
- \mu^2 \langle +-\rangle C^{\a\b}
\label{spinororth} 
\ee
using the anti-symmetrization $u_\a v_\b - u_\b v_\a = -u^\gamma 
v_\gamma C_{\a\b}$.  

Since we use 8 (real) components of $k_{(\pm)}^\b$ to describe 4 components 
of momentum, we have some freedom for restrictions.  A convenient restriction 
is
\be
\langle +-\rangle = [-+] = m
\label{light}
\ee
consistent with \rf{spinorline}.  This normalization eliminates a 
phase, however, within the inner products.  

Various normalizations of $\mu$
can be convenient.  For example, the orthonormal basis is
\be
\mu = {1\over\sqrt{\langle +- \rangle}} 
\qquad\Rightarrow\qquad
\psi_{(\pm)}^\b = {k_{(\pm)}^\b \over\sqrt{\langle + -  
\rangle}}, \qquad 
\psi_{(+)}^\a \psi_{(-)}^\b - \psi_{(-)}^\a \psi_{(+)}^\b =
- C^{\a\b}
\ee
However, for the above restriction on $\langle +- \rangle$, 
or in general since $|\langle +-\rangle|=m$, this choice is 
inconvenient for considering massless limits.  An alternative, used in 
conjunction with \rf{light}, is to drop the normalization factor:
\be
\mu = 1 \qquad\Rightarrow\qquad 
\psi_{(\pm)}^\b = k_{(\pm)}^\b, \qquad
\psi_{(+)}^\a \psi_{(-)}^\b - \psi_{(-)}^\a \psi_{(+)}^\b =
- m C^{\a\b}
\ee
In fact, these are the familiar choices $\bar uu = 1$ or $m$ for Dirac 
spinor line factors.

In the orthonormal case, the matrix $\psi_{(\pm)}^\b$ is just the 
Lorentz transformation from the rest frame to an arbitrary frame:  
Explicitly,
\be
k^{\a\da} = \psi_{(\b)}^\a (m\delta^{\b\gamma}) \bar\psi_{(\gamma)}^\da
\ee
The little group is SU(2) acting on the ``$(\a)$" indices, which leaves the 
momentum $k^{\a\da}$ invariant.  (Similar remarks apply when applied to 
the original fermionic actions.)  
This SU(2) invariance is just the freedom to change basis.
For the neutral (``Majorana") fermion, we note that the field equation 
$k_{\a\da} \psi^\a (k) + m\bar\psi_\da (-k) = 0$ relates $\psi_{(+)}$ for 
negative energy to $k_{(-)}$, etc.:
\be
\psi_{(\a)}^\a(-k_{(+)}-k_{(-)}) = 
\epsilon_{(\a)(\b)}\psi_{(\b)}^\a(k_{(+)}+k_{(-)})
\ee

Previous to \cite{Kife}, a more complicated choice of external line factors was 
used \cite{spinorhell}.  Both can be found by projecting with $\gamma\cdot k+m$ 
on a massless solution (for just $k_{(2)}$), but the newer choice projects on a 
Weyl spinor while the older one projects on a Majorana spinor, producing twice 
as many terms.  The two are related by an SU(2) little group transformation: 
explicitly,
\bqr 
\psi'^\gamma_{(\a)} = \Lambda_{(\a)}{}^{(\b)}\psi_{(\b)}^\gamma , \qquad
\Lambda_{(\a)}{}^{(\b)} = {1\over \sqrt 2}
\pmatrix{ 1 & -[+-]/m \cr  
 -\langle +- \rangle /m & 1 \cr}
\label{fermlines}
\fqr

This construction generalizes directly to arbitrary spin, 
since the use of this basis 
reduces any field equation to the rest frame.  
The result is always simple for the case of only undotted 
indices because there are no constraints 
(other than the Klein-Gordon equation) to satisfy.  
In contrast, the usual representations 
with mixed indices always require transversality conditions.

For example, the external line factors 
$L^{\a\b}$ for the massive vector particles have three components and represent 
the three independent polarizations of the massive vector field.  
Boosting from the rest 
frame to non-trivial momenta associated with each external line immediately yields 
\be
L_{(\a),(\b)}^{\gamma\delta} = \psi_{(\a)}^{(\gamma}\psi_{(\b)}^{\delta)} \ ,
\label{vectorline}
\ee 
with the same normalization as in the spinor case.  If we convert 
the $(\a)$ SU(2) indices to 3-vector notation with Pauli $\sigma$-matrices 
for $s_z=\pm 1,0$,
\be 
\sigma_{+,(\a)}{}^{(\b)} = \pmatrix{0 & 1 \cr 0 & 0}, \quad
\sigma_{-,(\a)}{}^{(\b)} = \pmatrix{0 & 0 \cr 1 & 0}, \quad
\sigma_{0,(\a)}{}^{(\b)} = {1\over\sqrt 2}\pmatrix{1 & 0 \cr 0 & -1}, 
\ee
we have, in the orthonormal basis, 
\bqr && 
L_+^{\a\b} = -i{k_{(+)}^\a k_{(+)}^\b\over \langle +- 
\rangle}, \quad 
L_-^{\a\b} = i {k_{(-)}^\a k_{(-)}^\b \over \langle +- 
\rangle}, 
\\ &&   
L_0^{\a\b} = -{i\over\sqrt{2} \langle +- 
\rangle} \bigl( k_{(+)}^\a  k_{(-)}^\b + k_{(-)}^\a k_{(+)}^\b \bigr) \ .
\label{dualpol} 
\fqr 
These polarizations satisfy the identities, 
\bqr  
L_+ \cdot L_- = 1 \qquad L_0 \cdot L_0 = 1 \qquad L_+\cdot L_0 
  = L_-\cdot L_0 = 0 \ , 
\fqr 
where $u\cdot v= u^{\a\b} v_{\a\b}$.  The orthogonality relation is
\bqr 
\sum_{\lambda=\pm,0} L_{\a\b}^\lambda L_{\gamma\delta}^{-\lambda} = 
 -{1\over 2}  C_{\gamma(\a} C_{\b)\delta} \ . 
\label{orthothree}
\fqr 
These relations are useful for cross-section calculations with and without 
explicit reference momenta choices for the external massive vectors.  

The three physical polarization vectors of the massive gauge field $A^{\a\da}$ 
satisfy 
\bqr 
k\cdot \varepsilon^\lambda = 0 \qquad  
 \varepsilon^\lambda(M,k) \cdot \varepsilon^{-\lambda}(M,k) 
 = 1 
\fqr 
and 
\bqr  
\sum_{\lambda = \pm,0} \varepsilon_{\a\da}^\lambda(M,k)  
 \varepsilon_{\b\db}^{\star,\lambda}(M,k) = C_{\a\b}C_{\da\db} +  
 {k_{\a\da} k_{\b\db}\over 2M^2} \ .  
\label{polid}
\fqr 
The 
set of three polarizations satisfying \rf{polid} are then also found 
by boosting with
$\psi_{(\a)}^\b$:
\bqr  
\varepsilon_+^{\a\da} = {k_{(+)}^\a k_{(-)}^\da \over M}, \quad 
\varepsilon_-^{\a\da} = {k_{(-)}^\a k_{(+)}^\da \over M}, \quad 
\varepsilon_0^{\a\da} = {1\over \sqrt 2 M} \bigl( k_{(+)}^\a 
 k_{(+)}^\da -  k_{(-)}^\a k_{(-)}^\da \bigr) \ . 
\fqr  
These results agree with those found by applying the 
relation $A^{\a\da}={i\over M} \partial^{\b\da}G_\b{}^\a$ to the $L$'s.

The use of these line factors together with our rules represents a
simplification over the standard Feynman rules using the polarization vectors
$\varepsilon_{\a\da}$ chosen for the massive vector lines.  In the latter case
the form of the four-component polarization vector is considerably
more complicated than the line factors above.  Use of these line factors 
is similar  to the use of spinor helicity for the massless vector bosons, 
in which the two polarization vectors are represented as bi-spinors,
\be 
\varepsilon^+_{\a\da}(k;q) = {q_\a k_{\da}\over q^\b k_\b}
\qquad
\varepsilon^-_{\a\da}(k;q) = -{k_\a q_{\da}\over
 q^{\db} k_{\db}}  \ .
\ee 
Different choices of $q$ in these representations lead to a shift 
\bqr  
\varepsilon_{\a\da}^\pm(k;q_1) - \varepsilon_{\a\da}^\pm(k;q_2) =  
 f(k;q_1,q_2) k_{\a\da} \ , 
\label{decoupling}
\fqr 
and the difference is zero on-shell as the longitudinal component 
decouples.  The adaptation of the spinor helicity formalism suitable
to describing the polarization vectors of the massive vector bosons 
follows as in the massive vector case.  Our approach generalizates 
this formalism to the dualized theory and the ambiguity is reflected 
in the decoupling of the transversal part of the massive gauge field.  

\subsection{Scattering}

To illustrate the use of the Feynman rules we compute in the dualized theory in 
(\ref{red}) the M\o ller and Compton scattering processes, i.e.\ 
$\psi \psi \rightarrow \chi \chi$ and $\psi G \rightarrow \chi G$.  

\begin{figure}
\begin{center}
\epsfig{file=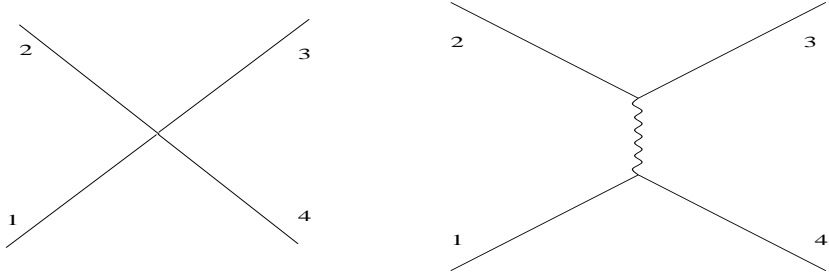,height=9cm,width=12cm}
\end{center}
\caption{The two contributions to M\"oller scattering, $\psi(k_1) \psi(k_2) 
\rightarrow 
\chi(k_3)\chi(k_4)$.}
\end{figure}

In M\o ller scattering we have the two diagrams shown 
in figure 1: (a) an intermediate massive photon line between two fermionic 
couplings and the permutation, and (b) the four-point fermionic vertex.
Labeling spin and momentum, we denote the initial states as
$\psi_{\a_1} (k_1)$ and $\psi_{\b_1}(k_2)$ and final states as $\chi_{\a_2}(k_3)$ 
and $\chi_{\b_2}(k_4)$.  The diagram (a) is
\bqr  
A_{\a_1 \b_1,\a_2 \b_2}(k_1,k_2;k_3,k_4) & = &
- {1\over M^2} \Bigl[ C_{\b_1\a_2} (k_3,k_4-k_1)_{\mu\nu} +  
 (k_1-k_4)^2 C_{\a_2\mu} C_{\nu\b_2} \Bigr]  \non 
& &  \times {1\over (k_1-k_4)^2 +M^2} 
 \Bigl[ C^{\mu\rho} C^{\nu\sigma} + C^{\mu\sigma} C^{\nu\rho} \Bigr] \non   
& & \times 
\Bigl[ C_{\a_1\b_2} (k_4,k_1-k_4)_{\rho\sigma} +  
 (k_1-k_4)^2 C_{\b_2\rho} C_{\sigma\a_1} \Bigr]   
\fqr 
where the latter terms in the first and second lines are symmetric due to 
the propagating vector in the intermediate state.  The expanded expression 
is 
\bqr 
A_{\a_1 \b_1,\a_2 \b_2}(k_1,k_2;k_3,k_4) & =  & 
-{1\over M^2} {1\over (k_1-k_4)^2+M^2} \times \non 
& & \Bigl\{  
 C_{\b_1\a_2} C_{\a_1\b_2} {\rm Tr} \bigl[ (k_4,k_1)(k_3,k_4) 
 - (k_4-k_1,k_3) (k_4,k_1) \bigr] \non 
& & - (k_1-k_4)^2 C_{\b_1\a_2} \bigl[ (k_3,k_4-k_1)_{\b_2\a_1} 
 + (k_3,k_4-k_1)_{\a_1\b_2} \bigr] \non 
& & - (k_1-k_4)^2 C_{\a_1\b_2} \bigl[ (k_4,k_1)_{\a_2\b_1} + (k_4,k_1)_{\b_1,a_2} 
 \bigr] \non 
& & - (k_1-k_4)^4 \bigl[ C_{\a_2\b_2} C_{\a_1\b_1} + C_{\a_2\a_1} C_{\b_2\b_1} 
 \bigr] \Bigr\}  \ , 
\fqr 
and the trace represents the contraction ${\rm Tr}(ab)=a^{\mu\nu} b_{\nu\mu}$. 
The second diagram (b) follows from the vertex in \rf{fourpointvertex}, and 
the amplitude is the sum of the two.  

Next we compare with the four fermion Fermi interaction theory 
\cite{Feynman:1958ty}.  
This is obtained by integrating out the massive vector, in the M\"oller process 
or by inspection of the dual Lagrangian in \rf{expanded}.  The dual theory has 
the nice feature that the intermediate vector exchange is an order higher in the 
fermion derivatives (at large values of $M$) than the $\psi^2\chi^2$ 
vertex interaction 
that is obtained directly in the dual theory in \rf{red} (the dual theory is more 
natural in this regard with the V$-$A theory).  The dual Fermi interactions are 
modeled by 
\bqr  
{\cal L}_{\rm Fermi} & = &  -\psi^\a \left( \Box-m^2 \right) \chi_\a  
-{1\over 2M^2} \left( \psi^\b\partial_{\a\da} \chi_\b\right) 
 \left(\psi^\rho \partial^{\a\da}\chi_\rho\right)  
\non & &  
-{1\over 2M^2} \psi^\b\partial_{\a\da}\chi_\b 
\partial^{\rho\da} \left[ \psi^{(\a}\chi_{\rho)} \right] 
+{1\over 8M^2} \psi^{(\a} \chi^{\b)} \Box \left( \psi_{(\a} 
\chi_{\b)}\right)  \ . 
\label{dualfermiint}
\fqr 
The form in \rf{dualfermiint} is in a second order fermionic form; we can 
undualize it to write it in first order form via, 
\bq  
{\cal L}_{\rm Fermi} =  
- {\bar\psi}^{\da} i\pa_{\a\da} \psi^\a 
- {\bar\chi}^{\da} i\pa_{\a\da} \chi^{\a} +
 m(\psi^\a\chi_\a  + {\bar\psi}^{\da}{\bar\chi}_{\da} ) 
+ {1\over M^2} \psi^\a {\bar\psi}^\da \chi^\b {\bar\chi}^\db 
 {\cal O}_{\a\da,\b\db} 
 \ . 
\label{fermiint} 
\fq 
Integrating half of the fermionic components in \rf{fermiint} generates the theory 
in \rf{dualfermiint} after scaling the fermionic fields by $\psi\rightarrow m\psi$, 
$\chi\rightarrow m\chi$ and keeping terms up to the quadratic order in derivatives.  
The matrix ${\cal O}_{\a\da,\b\db}$ is determined in the process.  Decoupling the 
undualized massive vector in the M\"oller scattering induces the matrix 
${\cal O}_{\a\da,\b\db}=C_{\a\b} C_{\da\db}$.  

For Compton scattering $\psi G \rightarrow \chi G$ there are three diagrams: 
1) two with an intermediate photon line via the vertices in (\ref{ruleone}), 
2) one from the four-point vertex in (\ref{ruletwo}).  We denote the 
momenta of the fermions by $k_1$ and $k_2$ and those of the photons 
$p_1$ and $p_2$.  The indices of the fermions are $\a_1$ and 
$\a_2$ while those of the photons are $\mu_j,\nu_j$.  
The first diagram gives 
\bqr  
& & 
A_{1,\a_1\a_2\mu_1\nu_1\mu_2\nu_2} = \left({1\over M^2}\right) 
\Bigl[ -C_{\a_1\a_2} (k_1+p_1,p_1)_{\mu_1\nu_1} (k_2p_2)_{\mu_2\nu_2} 
\non & & \hskip .2in  
-{M^2\over 2} C_{\a_2(\mu_1} C_{\nu_1)\a_1} (k_2p_2)_{\mu_2\nu_2}  
+ {M^2\over 2} (k_1+p_1,p_1)_{\mu_1\nu_1} C_{\a_2(\mu_2} C_{\nu_2)\a_1} 
\non & & \hskip .2in 
+ {M^4\over 4} C_{\a_2(\mu_2} C_{\nu_2)(\mu_1} C_{\nu_1)\a_1} \Bigr] 
{1\over (k_1+p_1)^2+m^2}   
\, 
\label{diagramone} 
\fqr  
together with the crossed diagram by symmetrizing with respect 
to the two photon legs, 
\bqr    
& & 
A_{2,\a_1\a_2\mu_1\nu_1\mu_2\nu_2} = \left({1\over M^2}\right) \Bigl[ 
-C_{\a_1\a_2} (k_1+p_2,p_2)_{\mu_2\nu_2} (k_2p_1)_{\mu_1\nu_1} 
\non & & \hskip .2in 
-{M^2\over 2} C_{\a_2(\mu_2} C_{\nu_2)\a_1} (k_2p_1)_{\mu_1\nu_1} 
+{M^2\over 2} (k_1+p_2,p_2)_{\mu_2\nu_2} C_{\a_2(\mu_1} C_{\nu_1)\a_1}  
\non & & \hskip .2in 
+{M^4\over 4} C_{\a_2(\mu_1} C_{\nu_1)(\mu_2} C_{\nu_2)\a_1} \Bigr] 
{1\over (k_1+p_2)^2+m^2} 
\ .
\label{diagramtwo} 
\fqr  
The four-point vertex for $\psi\chi G^2$ is identical to that 
in (\ref{ruletwo}) and generates the final diagram to this process.   

Squaring the amplitude to obtain the cross-section, via summing over 
final states and averaging over initial ones, involves the identities 
and \rf{orthothree} for the external fermion and vector line factors, 
respectively.   We complete the amplitude calculations for the Compton 
process in the remainder of this section, taking the fermionic matter 
to be massless.  In the massless limit the line factors associated with 
the second order fermions \cite{Morgan:1995te}, 
\bqr  
\epsilon^+_\a (k) = k_\a \qquad \epsilon^-_\a(k) = {q_\a\over 
\langle qk\rangle}  
\fqr    
are used in the calculation.  The reference momenta and first line factor 
choices are  
\bqr 
\psi_\a(k_1) = k_{1+,\a} \qquad \chi_{\a}(k_2) = 
 { k_{1+,\a}\over \langle 1+2+\rangle }
\fqr 
and 
\bqr  
L_{1,\mu\nu}(p_1) = -i {k_{1+,\mu}k_{1+,\nu}\over \langle 1+1-\rangle} \qquad 
 p_1=k_{1+}+k_{1-} 
\fqr 
\bqr  
L_{2,\mu\nu}(p_2) = i {k_{2-,\mu}k_{2-,\nu}\over \langle 2+2-\rangle} \qquad 
 p_2=k_{2+}+k_{2-} \ ,  
\fqr 
corresponding to the scattering process $\psi G^+ \rightarrow \chi G^-$.  
With these choices all terms from the first diagram and third diagram are 
immediately equal to zero; furthermore we have the normalizations $\langle j+ 
j-\rangle=[j-j+]=M$.  Only the second term in the second diagram is 
non-vanishing and it generates, 
\bqr  
A(+,-)= {\langle 1+2-\rangle^2 [2+1-] \over \langle 2+2-\rangle} 
 {1\over (k_{1+}+p_2)^2} \ .   
\fqr 

The opposite helicity configuration for the gauge bosons, parameterized by 
\bqr 
L_{1,\mu\nu}(p_1) = i {k_{1-,\mu}k_{1-,\nu}\over \langle 1+1-\rangle} \qquad 
 p_1=k_{1+}+k_{1-} 
\fqr 
\bqr  
L_{2,\mu\nu}(p_2) = -i {k_{2+,\mu}k_{2+,\nu}\over \langle 2+2-\rangle} \qquad 
 p_2=k_{2+}+k_{2-} \ ,  
\fqr 
receives a non-vanishing contribution from the fourth term in the first 
diagram and the first term in the second diagram equals zero.  The sum 
of the contributing terms equals 
\bqr  
A(-,+) &=& {M\over (k_{1+}+p_1)^2} \langle 1-2+\rangle 
 \non & & \hskip -.8in  + 
{1\over (k_{1-}+p_2)^2} \Bigl[ (k_{1+}+k_{2+})^2 {\langle 1-1+\rangle\over M} 
 + M ([1+2-] + \langle 1-2+\rangle)\Bigr] \ ,  
\fqr 
in terms of the reference momenta.  The two processes $\psi G^\pm\rightarrow 
\chi G^\pm$ are trivially zero after contractions.  Squaring the sum of these 
amplitudes and taking the high-energy limit, $M\rightarrow 0$, generates  
\bqr 
\sum \vert M\vert^2 \rightarrow {16\over M^2} 
 {(p_1\cdot k_{2+})^2 \over k_{1+}\cdot p_2}  \ . 
\fqr 
Normalizing by the line factor associated with the fermion $k_{1+}\cdot k_{2+} 
=(k_{1+}+k_{2-})\cdot k_{2+} -M^2 \rightarrow p_2\cdot k_{2+}$ gives the known 
result,  
\bqr 
\sum \vert {\tilde M}\vert^2 \rightarrow 
{16\over M^2} ~p_2\cdot k_{1+} p_2\cdot k_{2+} \ .  
\fqr 

\section{Non-abelian dualizations}
\subsection{Vectors} 

In this section we shall consider a general spontaneously broken 
non-abelian gauge theory \cite{Chalmers:1999jb}; the scalar and fermion 
matter content is taken to be generic.  The dualized form to this
theory is obtained by expressing the original minimally coupled
theory in first order form using a Lagrange multipler field strength
$G_{\a\b}$:
\be 
{\cal L} = {\rm Tr}~ -{g^2\over 2} G^{\a\b}G_{\a\b} +
G^{\a\b}F_{\a\b}  + \nabla^{\a\da}\phi^\dagger
\nabla_{\a\da}\phi
- {\bar\psi}^{\da}i\nabla_{\a\da} \psi^{\a} + V(\phi)
\ .
\label{lagrangian}
\ee 
Integrating out the dualized gauge fields $G^{\a\b}$ gives back the 
usual field theory formulation.  However, this first order form is 
now quadratic in the gauge field $A_{\a\db}$; we shall 
rather functionally integrate out $A^{\a\db}$ \cite{Chalmers:1997sg} 
and arrive at a theory defined in terms of $G_{\a\b}$. 

Integrating out the gauge connection for those vectors acquiring a
mass through spontaneous symmetry breaking gives rise to the
T-dualized Lagrangian examined in \cite{Chalmers:1997sg}.  Prior to gauge
fixing we obtain the theory after integrating out
$A_{\a\da}$,
\be
{\cal L}^d  = {\rm Tr}~ -{1\over 2} G^{\a\b}G_{\a\b} +
  V_{\a,a}{}^{\dt\gamma} \left[ X^{-1}\right]_{ab}^{\a\b}
V_{\b\dt\gamma,b}
- {\bar\psi}^{\da,a} i\partial_{\a\da} \psi^\a_a
 + \partial^{\a\da}\phi^{\ast,a} \partial_{\a\da}\phi_a
 + V(\phi) \ .
\label{nondualform}
\ee
The remaining massless vectors not Higgsed by the scalar interactions and 
in (\ref{nondualform}) not indexed by $a$, give the standard
(undualized) contribution to the  Lagrangian:
\be
{\cal L}^u = {\rm Tr}~~{1\over 2} F^{\a\b}F_{\a\b}
+ (\nabla^{\a\da}\phi)^\dagger \nabla_{\a\da}\phi
- {\bar\psi}^{\da}i\nabla_{\a\da} \psi^{\a} \ .
\ee
The full theory in eq.\ (\ref{lagrangian}) is ${\cal L}_d+{\cal L}_u$.  The
gauge-fixing may be performed before the integration of the gauge
field; in unitary gauge this would eliminate one component  of the
complex scalar.  We will do so in the next section relevant to the dual
of the electroweak model.  In the above, the vector $V$ and matrix $X$
in eq.~(\ref{nondualform}) are found to be
\be 
V_a^{\a\da}=\partial_\gamma{}^{\da} G_a^{\a\gamma}
+ i {\bar\psi}^{\da} T_a^{\psi} \psi^\a + \phi^\dagger T_a^{\phi}
 \left(\partial^{\a\da} \phi\right)
 - \left(\partial^{\a\da} \phi^\dagger\right) T_a^{\phi}
 \phi
\ ,
\ee
and
\be 
X_{ab}{}^{\a\b} =  if_{ab}{}^c G_c{}^{\a\b} +
 \phi^\dagger \left\{ T_a^{\phi}, T_b^{\phi} \right\} \phi ~C^{\a\b}
 \ .
\ee
We need to expand the form in eq.~(\ref{nondualform}) about the broken
phase in order to generate the Feynman rules, of which there are an
infinite  number arising from the Taylor expansion about the vacuum
value of the  scalar field.  We write the expansion as
$\phi={\tilde\phi}+v$, and of course, the lowest order term in
eq.~(\ref{nondualform}) is kept by keeping only the quadratic in $v$
term in the inverse of the matrix $X^{\a\b}_{ab}$, which
generates the  mass term for the vector bosons.  These rules will be
derived in the following sections.

\subsection{Spinors}

Our aim is to eliminate to a large extent the dotted $\da$ index 
from appearing in the gauge theory coupled to matter of various spin content.  
We will see that this allows for a ``color ordering'' for internal spin 
structures which leads to a further minimization of amplitudes into smaller 
gauge invariant subsets; the minimization of amplitudes into subsets is 
advantageous because calculations of smaller gauge-invariant subsets of 
amplitudes is simpler and there are relations between these subsets (as 
in the case of subleading in color contributions to partial amplitudes: 
these subleading trace structure partial amplitudes $A_{n;m}$ can be 
obtained from $A_{n;1}$).  This is analogous to the effect that spinor 
helicity has in minimizing the redundancy of the gauge invariant interactions 
and is similar to the color ordered breakup of amplitudes.  

We demonstrate here then the dualization of the non-abelian model
in the previous section combined with the reduced fermion Lagrangian
derived in \cite{Chalmers:1999ui}.  Recall that the fermion reduction is derived by
integrating out half  of the fermionic components contained in the Lagrangian 
in eq.(10).  

Upon using the field equation for ${\bar\psi}$, we eliminate this
field and obtain the fermionic contribution to the gauge theory in 
(\ref{reduce}); the $\Box$ is gauge covariantized, i.e.\ 
$2\Box=\nabla^{\a\da} \nabla_{\a\da} = 
(\partial^{\a\da}+ iA^{\a\da}) 
(\partial_{\a\da}+ iA_{\a\da})$, and 
generates tri-linear and quartic couplings.  The fully dualized matter 
theory is found through (\ref{lagrangian}), but together with the 
fermion contribution in (\ref{reduce}) (which is quadratic in 
$A_{\a\da}$).   In the following we
combine this integrating out procedure with the dualization of the
vector fields.

To generate the dualized system of the non-abelian gauge theory 
coupled to fermions we introduce as before a first order form through
\be 
{\cal L}_g = {\rm Tr}~ -{1\over 2} G^{\a\b}G_{\a\b} +
G^{\a\b}F_{\a\b} \ .
\ee
We proceed by integrating out the massive vector fields as in the
previous section, after including the Lagrangians in eqs. (\ref{lagrangian}) 
and (\ref{nondualform}).  The general form of the dual Lagrangian 
taking into account the reduced fermion contribution in eq.~(\ref{reduce}) 
is then (after integrating out $A^{\a\da}$), 
\bqr  
{\cal L}^d  &=& {\rm Tr}(-{1\over 2} G^{\a\b}G_{\a\b}) +
 V_{\a,a}{}^{\dt\gamma}
\left[ X^{-1}\right]_{ab}^{\a\b}  V_{\b\dt\gamma,b}
\non & &  + ~ \partial^{\a\da} {\psi}^{\b,a} \partial_{\a\da}
\psi_{a,\b}
 + \partial^{\a\da}\phi^{\ast,a} \partial_{\a\da}\phi_a
 + V(\phi) \ .
\label{dualfermform}
\fqr 
The vector $V_a^{\a\da}$ is given by
\be
V_a^{\a\da}= \partial_\gamma{}^{\da} \left(
G_a^{\a\gamma}
+ \psi^\a T_a^{\psi} \psi^\gamma\right)
+ \psi^\gamma T_a^{\psi} \on{\leftrightarrow}{\partial}{}^{\a\da}
\psi_\gamma
+ \phi^\dagger T_a^{\phi} \on{\leftrightarrow}{\partial}{}^{\a\da}
\phi
\ , 
\label{genvector}
\ee
and the matrix $X_{ab}{}^{\a\b}$ is
\be 
X_{ab}{}^{\a\b} =  if_{ab}{}^c G_c{}^{\a\b} +
 \left(\phi^\dagger \left\{ T_a^{\phi}, T_b^{\phi} \right\} \phi
+ \psi^\gamma \left\{T_a^{\psi}, T_b^{\psi} \right\} \psi_\gamma
\right) C^{\a\b}
 \ .
\ee 
Note that the resulting action is complex.  We have also left the 
representation content of the matter under the gauge group arbitrary 
$(T_a^\phi,T_a^\psi)$; we shall consider specific representations in 
the following subsections.  

The form (\ref{dualfermform}) of the dualized theory may be 
found by comparison with
the first order Lagrangian in  eq.~(\ref{lagrangian}); the
two couplings in eq.~(\ref{reduce}), the scalar-like box and the
selfdual tensor $F_{\a\b}$, couple as the
$G^{\a\b}F_{\a\b}$ term and as the scalar $\phi\Box\phi$ ones.
Note that in distinction to the result in eq.~(1.2), the fermions
enter into the denominator of the dualized theory through the matrix
$X_{ab}^{\a\b}$; the entire contribution for $X^{-1}$ must be 
expanded about the vacuum value leading to an infinite number of 
interaction terms (but a finite number at given order in perturbation 
theory).  In this reduced dualized theory we have chirally
minimized the theory in the sense that all of the fields are labelled by only
one type of spin index, i.e.\ $G^{\a\b}$, $\psi^\a$, and $\phi$.
A single (for fermions) and double (for gauge fields) line formulation 
naturally follows associated with the contractions of the $\a$-type 
index. 

\section{Feynman Rules}

In this section we will derive the Feynman rules to the reduced dualized
theory of the previous section.  (In section 6, we will apply these to the example 
of the four-point vector amplitude.)  We also make explicit the double 
line representation for the Feynman graphs in accord with the spin ordering.

To find the couplings to fourth order in the gauge fields, we need
to expand the inverse to the matrix $X^{\a\b}_{ij}$ to second order
about
its background values of the scalar fields.  The matrix $X$ has the form,
\be
X_{ij}^{\a\b}=G_k^{\a\b} M^k_{ij} + Q_{ij}C^{\a\b} \ .
\label{Xmatrixone}
\ee
The first term in eq.~(\ref{Xmatrixone}) is symmetric in $(\a,\b)$ and
anti-symmetric in $(i,j)$, and the second one is anti-symmetric in
$(\a,\b)$ and symmetric  in $(i,j)$.  

The perturbative form of the inverse to the matrix $X$ is obtained
from the expansion
\be
\Bigl[ {1\over X} \Bigr]_{\a}^{ij,\b} =
 {1\over G_{(m)} M^{(m)} + Q \delta} =
 {1\over Q} \Bigl[{1\over \delta +Q^{-1} G_{(m)} M^{(m)} } \Bigr]
\ ,
\ee
around the mass matrix $Q_{ij}=\phi^\dagger\{ T_i^\phi,T_j^\phi\}\phi$.  
Expanding $X=\langle X\rangle + \triangle X$ is not direct because there 
is a field dependent coefficient multiplying the $\langle X\rangle$ term.  
A first order expansion generates Feynman rules to third order
in the $G$-fields,
\bqr 
\Bigl[ {1\over X} \Bigr]_{\a}^{ij,\b} & = &
\bigl[ Q^{-1}_{ik} \bigr]
 \bigl[ \delta_\a{}^\b \delta_{kj} -
 Q^{-1}_{kl}~ G_{(a)\a}{}^\b M^{(a)}_{lj} + \ldots \bigr]
\non & = &  
 \delta_\a{}^\b Q^{-1}_{ij} +
 Q^{-1}_{ik} Q^{-1}_{kl} G_\a{}^{\b(a)} M^{(a)}_{lj}
+ \ldots \ .
\label{firstorder}
\fqr 
Higher order terms to the inverse matrix are obtained by performing 
the Taylor expansion and contain more powers of the matrix 
$G_{(k)}^{\a\b} M^{(k)}_{ij}$.  Including the fermion 
contributions, for example, gives the lowest order expansion: 
\be
\Bigl[ {1\over X} \Bigr]_{\a}^{ij,\b}
 = \delta_\a{}^\b Q^{-1}_{ij} +
 Q^{-1}_{ik} Q^{-1}_{kl}
 \Bigl[ G_{\a}{}^{\b(a)} M^{(a)}_{lj}  +
\delta_\a{}^\b \psi^\gamma\left\{ T_l^\psi, T_j^\psi \right\}
\psi_\gamma  \Bigr]
+ \ldots \ ,
\label{reducedfirstorder}
\ee
with the $Q_{ij}$ matrix the same form, i.e.\ $Q_{ij}=\phi^\dagger
\left\{ T_i^\phi, T_j^\phi \right\} \phi$.  Although we may write a 
general form for the perturbative inverse, we shall not do so here 
as it is not relevant to the following discussion.  These expansions 
suffice to generate the quartic couplings of the gauge fields. 

We now describe how the group theory color ordering combines with 
the spin ordering.  The mass matrix for the scalars in the fundamental 
representation has the particularly simple form in an $SU(2)$ 
gauge theory, 
\be
Q_{ab} =\phi^\dagger \left\{ T_a^\phi, T_b^\phi\right\} \phi =
 {1\over 2}\delta_{ab} ~\phi^\dagger\phi \ .
\ee
This form of the mass matrix $Q_{ij}$ simplifies the derivation of the
Feynman rules in this section because its inverse is diagonal.  On the
other hand, the general fundamental representation of $SU(N)$
satisfies the identity,
\be
\left\{ T_a^\phi, T_b^\phi\right\} = {1\over N}\delta_{ab}
+ d_{abc} T^c \ ,
\ee
and the mass matrix is only symmetric.  Along the directions 
$d_{abc} v^\dagger T^c v=0$, the inverse is also quite simple.  The 
inverse for general representations incorrectly prohibits factoring 
out of the group theory via color ordering because we need to 
determine the appropriate color flow along the propagators.  The 
color flow with the metric generated for the propagators above 
in the general case is provided in a general representation of the 
matter content.  

The fourth order in $G_{\a\b}$ Feynman rules require a second order
expansion of $1/X_{\a\b}^{ab}$.   We first define
the inverse matrix to the background values of the mass matrix by
\be
I_{ac} ~v^\dagger \left\{ T_c^\phi, T_b^\phi\right\} v = \delta_{ab} \ .
\label{inversematrix}
\ee
We shall drop the couplings to scalars; the theory is
\bqr 
{\cal L}_r &=& \Bigl[ \partial_\gamma{}^{\da} 
 \left( G_a^{\a\gamma}
+ \psi^\a T_a^{\psi} \psi^\gamma \right)
+ \psi^\b T_a^{\psi}
 \on{\leftrightarrow}{\partial}{}^{\a\da} \psi_\b
\Bigr]
\left[ {1\over X}\right]_{ab}^{\a\b}
\non & &
\times \Bigl[ \partial_{\rho\da} \left( G_b^{\b\rho}
+ \psi^\b T_b^{\psi} \psi^\rho \right)
+ \psi^\mu T_b^{\psi} \on{\leftrightarrow}{\partial}{}^\b{}_{\da}
 \psi_\mu
 \Bigr] \ ,
\label{thirdord}
\fqr 
where to second order we have
\bqr 
\left[ {1\over X}\right]^{\a\b}_{ab} &=&
  {1\over (v^\dagger v)} C^{\a\b} I_{ab}
 + {1\over (v^\dagger v)^2} I_{ac} I_{cd} \left( G^{\a\b}_n
M^{(n)}_{db}
 + C^{\a\b} \psi^\gamma \{ T_d^\psi, T_b^\psi\} \psi_\gamma \right)
\non & &  
+{1\over 8(v^\dagger v)^3} I_{ac} I_{cd} \left(
 G^{\a\delta}_n M^{(n)}_{de} + C^{\a\delta}
 \psi^\gamma \{ T_d^\psi, T_e^\psi\} \psi_\gamma \right)
\non & & \qquad\qquad 
\times  
I_{ef}\left( G_\delta{}^\b_n M^{(n)}_{fb} + \delta_\delta{}^\b
 \psi^\gamma \{ T_f^\psi, T_b^\psi\} \psi_\gamma \right) \ .
\fqr  
The metric $I_{ab}$ is the metric used to contract the color indices 
arising at the vertices.  The net color flow, even for general $I_{ab}$, 
may be factored out as in the usual formulation of Yang-Mills theory.  

This expansion of the inverse matrix may be thought of as one in the
dimension of the operators,  or inversely in the counting of the Higgs
vacuum value which enters through the background value of the mass
matrix.  The color ordered Feynman rules are simplest when $I_{ab}$ 
is diagonal (e.g.\ $SU(2)$ fundamental).

The vertices in (\ref{thirdord}) possess two momenta with a
string of Kronecker deltas absorbing the indices; this suggests a
simple double line representation associated with the $\a$
indices  to the Feynman rules similar to that of color flow (with
Chan-Paton factors).    In the diagrams we represent a Kronecker delta
symbol connecting the indices at either end by a single line.  We have
to also represent the product $k_{i\a\da}
k_j^{\b\da}$ from the contraction of  two momenta in the
vertices; a dot on the line indicates that the matrix
$k_{i\a\da} k_j^{\b\da}$ is used to soak up the
indices on the lines $i$ and $j$.  The general  fermion or gauge boson
(i.e.\ $G^{\a\b}$) line possesses these  contractions, as can be
seen in $V_a^{\a\da}$ in eq.\ (\ref{genvector}).

\begin{figure}[ht]
\begin{center}
\epsfig{file=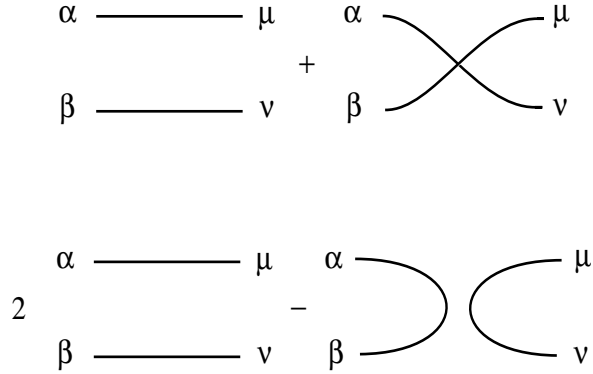,height=5cm,width=8cm}
\end{center}
\caption{The two terms contributing to the propagator.}
\end{figure}
 
\section{Spin Ordering} 

We now examine in detail the lower-point vertices and describe the 
spin ordering in practice.  The lowest order expansion generates the 
propagator in the form,
\be 
{\cal L}_{G^2} = {I_{ab}\over 2 M^2} ~G^{\a\b,a}
 \left[\Box - M^2 \right] G_{\a\b}^b  \ , 
\ee  
with $I_{ab}$ defined in \rf{inversematrix}.  We rescale $G^{\a\b} 
\rightarrow MG^{\a\b}$ here and in the following to simplify
the form of the propagator and vertices.  In momentum space the
color ordered propagator has the two equivalent tensor structures
\bqr 
\langle G_\a{}^\b (k) G_{\mu}{}^\nu(-k)\rangle & = &
 {1\over k^2 + M^2} \left[ \delta_\a^\nu \delta^\b_\mu +
 \delta_{\a\mu} \delta^{\b\nu} \right]
\non & = &  
{1\over k^2+M^2} \left[ 2 \delta_\a^\nu \delta^\b_\mu  -
\delta_\a^\b \delta_\mu^\nu \right]  \ .
\fqr 

The group theory factor associated with the propagator is contained
in the color factor $I_{ab}$.  The forms of the propagators are
illustrated within the double line formulation in figure 2.  In 
figure 2 we have not included the color flow, but represent the lines 
as contractions of the $\a,\b$-type vertices.

The three- and four-point couplings containing only the
$G$-fields are obtained from the interactions
\be
{\cal L}_{G^3} = {1\over M} ~
 \bigl(\partial_\gamma{}^{\da} G_a^{\a\gamma}\bigr)
G_{b,\a\b}
 \bigl( \partial_{\rho\da}  G_c^{\b\rho} \bigr) \ , 
\ee
\be
{\cal L}_{G^4}= {1\over M^2}  ~
\bigl( \partial_\gamma{}^{\da} G_a^{\a\gamma} \bigr) ~
 G_{b,\a}{}^\b G_{c,\b}{}^\delta
 \bigl( \partial_{\rho\da} G_{d,\delta}{}^\rho \bigr)  \ ,
\label{couplings}
\ee 
where the tensors $I_a{}^d I_d{}^e f^{bc}{}_e$ and $I_{ae}I^{ei} f^b{}_{ig} 
I_{gh} f_{hb}{}^c$ represent the color structure and have been removed 
in the above; they are obvious from the graphical interpretation.  
We shall also use the double line representation to graphically represent the
vertices; unlike the propagator, we must use a line with a dot to represent
the momenta contraction on the lines.

The three-point vertex found from ${\cal L}_{G^3}$ contains the terms
\bqr
\langle G^{\a_1}_{a_1}{}_{\b_1}(k_1)  
G^{\a_2}_{a_2}{}_{\b_2} (k_2) 
  G^{\a_3}_{a_3}{}_{\b_3} (k_3) \rangle 
& = & -{1\over M} \Bigl(
 I^{a_1d}I_d{}^e f^{a_2a_3}{}_e  ~ (12)^{\a_1}{}_{\b_2} ~ 
 \delta^{\a_2}_{\b_3} \delta^{\a_3}_{\b_1} \nonumber \\
&& + I^{a_2d}I_d{}^e f^{a_3a_1}{}_e ~(23)^{\a_2}{}_{\b_3}
~\delta^{\a_3}_{\b_1} \delta^{\a_1}_{\b_2} \nonumber \\
&& + I^{a_3d}I_d{}^e f^{a_1a_2}{}_e ~(31)^{\a_3}{}_{\b_1}
~\delta^{\a_1}_{\b_2} \delta^{\a_2}_{\b_3} \Bigr) \ ,
\label{threepoint}
\fqr 
where we have explicitly summed over the ordering of the external labels
to make the rule Bose symmetric.  In addition we have to symmetrize in the
the pairs of indices $(\a_1,\b_1)$, $(\a_2,\b_2)$, and
$(\a_3,\b_3)$; we do not include these terms because
the propagator and external line factors explicitly twist these indices and
symmetrizes the pairs.  From now on we shall drop the color matrices and 
leave the result color-ordered with the exception of the following four-point 
example; furthermore we use the notation, 
\be
(ij)^{\a\b} = k_i{}^{\a\da} k_j^\b{}_{\da}   
\ee
to define the often recurring matrix from the vertex algebra.

\begin{figure}
\begin{center}
\epsfig{file=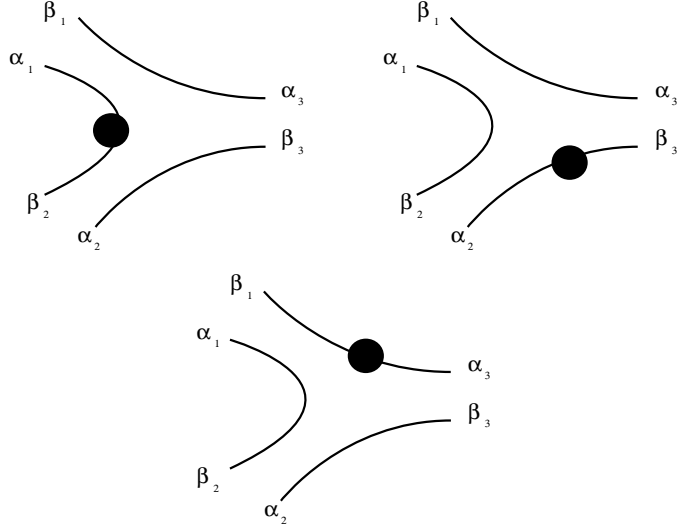,height=7cm,width=9cm}
\end{center}
\caption{The three terms of the color-ordered three-point vertex.}
\end{figure}

The color-ordered three-point vertex is found by taking the $(123)$ ordering
of the group trace structures represented by the Levi-Cevita tensors,
\be
f^{a_1a_2a_3}={\rm Tr}\left( \left[ T^{a_1},T^{a_2}\right] T^{a_3}
\right) \ .
\ee
In eliminating the group theory factors we find the color
ordered form of the three-point vertex to be
\be
V_{123}{}^{\a_1\a_2\a_3}_{\b_1\b_2\b_3} =
{1\over M} \Bigl( (12)^{\a_1}{}_{\b_2}
~\delta^{\a_2}_{\b_3}  \delta^{\a_3}_{\b_1}
+ (23)^{\a_2}{}_{\b_3}
~\delta^{\a_3}_{\b_1} \delta^{\a_1}_{\b_2}
+ (31)^{\a_3}{}_{\b_1}
~\delta^{\a_1}_{\b_2} \delta^{\a_2}_{\b_3} \Bigr) \ .
\label{colorordthree}
\ee  

This vertex is illustrated graphically in figure 3.  The dot in figure 
3 denotes the contraction of the momenta as found in the vertex in
eq.~(\ref{colorordthree}) from the factors $(ij)$.  

The momentum space four-point vertex is derived in a similar fashion.  We
label all of the lines with the pairs of indices $(\a_i,\b_i)$; its
Bose symmetric form produces the terms in the case of $SU(2)$,
\bqr 
&& \langle  G^{\a_1\b_1}_{a_1} (k_1)  G^{\a_2\b_2}_{a_2}(k_2)
  G^{\a_3\b_3}_{a_3}(k_3) G^{\a_4\b_4}_{a_4}(k_4) \rangle =
\non & &
\quad
 {1\over M^2} \Bigl( \epsilon^{a_1a_2l}\epsilon_l{}^{a_3a_4} ~
 (12)^{\a_1}{}_{\b_2} ~
\delta^{\a_2}_{\b_3} \delta^{\a_3}_{\b_4}
\delta^{\a_4}_{\b_1}
+ \epsilon^{a_4a_1l}\epsilon_l{}^{a_3a_2} ~
 (41)^{\a_4}{}_{\b_1} ~
\delta^{\a_1}_{\b_2} \delta^{\a_2}_{\b_3}
\delta^{\a_3}_{\b_4}
\non & &
+  \epsilon^{a_2a_3l}\epsilon_l{}^{a_1a_4}  ~
  (23)^{\a}{}_{\b_3} ~
\delta^{\a_3}_{\b_4} \delta^{\a_4}_{\b_1}
\delta^{\a_1}_{\b_2}
+  \epsilon^{a_3a_4l}\epsilon_l{}^{a_2a_1} ~
  (34)^{\a_3}{}_{\b_4} ~
\delta^{\a_4}_{\b_1} \delta^{\a_1}_{\b_2}
\delta^{\a_2}_{\b_3}
\non & &
+ \epsilon^{a_1a_2l}\epsilon_l{}^{a_3a_4}~
 (12)^{\a_1}{}_{\b_2} ~
\delta^{\a_2}_{\b_4} \delta^{\a_4}_{\b_3}
\delta^{\a_4}_{\b_1}
+ \epsilon^{a_4a_1l}\epsilon_l{}^{a_2a_3}~
 (41)^{\a_4}{}_{\b_1}~
\delta^{\a_1}_{\b_3} \delta^{\a_3}_{\b_2}
\delta^{\a_2}_{\b_4}
\non & &
+  \epsilon^{a_2a_3l}\epsilon_l{}^{a_4a_1} ~
 (23)^{\a_2}{}_{\b_3} ~
\delta^{\a_4}_{\b_2} \delta^{\a_1}_{\b_4}
\delta^{\a_3}_{\b_1}
+ \epsilon^{a_3a_4l}\epsilon_l{}^{a_1a_2} ~
 (34)^{\a_3}{}_{\b_4} ~
\delta^{\a_4}_{\b_2} \delta^{\a_1}_{\b_3}
\delta^{\a_2}_{\b_1} + \ldots \Bigr) 
 \ .
\label{fourpoint}
\eeqs
There are in addition terms not included in the symmetric sum which do
not possess a possible ordering $\epsilon^{a_1 a_2m}\epsilon_m{}^{a_3
a_4}$ or $\epsilon^{a_2 a_3m}\epsilon_m{}^{a_4 a_1}$ of the group
theory factors. (The contractions of the epsilon tensors can be written as 
products of Kronecker delta functions, but in a color ordering the group 
theory factorizes directly from the kinematics.)  The eight terms above 
in eq.~(\ref{fourpoint}) must also be twisted within the pairs $(\a_i,\b_i)$ 
to generate the complete four-point vertex; however, as with the three-point 
vertex the propagator and external line factors explicitly perform the
twisting of these indices.  We may strip the color from the above vertex 
by using 
\be
\epsilon^{a_1a_2l}\epsilon_l{}^{a_3a_4} =
 {\rm Tr}\left[ T^{a_1},T^{a_2}\right] \left[ T^{a_3},T^{a_4}\right] \ ,
\ee
and collecting all contributions to the trace structure with the
ordering $(1234)$. 

In general, for $SU(N)$, we need the metric $I_{ab}$ to join the 
trace generators at each external line.  Doing so gives the color-ordered 
form the same as in the $SU(2)$ example, 
\bqr 
V_{1234}{}^{\a_1\a_2\a_3\a_4}_{\b_1\b_2\b_3\b_4}
 & = &   {1\over M^2} \Bigl( 
 (12)^{\a_1}{}_{\b_2} ~
\delta^{\a_2}_{\b_3} \delta^{\a_3}_{\b_4}
\delta^{\a_4}_{\b_1}
-  (41)^{\a_4}{}_{\b_1}~
\delta^{\a_1}_{\b_2} \delta^{\a_2}_{\b_3}
\delta^{\a_3}_{\b_4}
\non & &
-  (23)^{\a_2}{}_{\b_3} ~
\delta^{\a_3}_{\b_4} \delta^{\a_4}_{\b_1}
\delta^{\a_1}_{\b_2}
- (34)^{\a_3}{}_{\b_4} ~
\delta^{\a_4}_{\b_1} \delta^{\a_1}_{\b_2}
\delta^{\a_2}_{\b_3}
\non & &
+  (12)^{\a_1}{}_{\b_2} ~
\delta^{\a_2}_{\b_4} \delta^{\a_4}_{\b_3}
\delta^{\a_3}_{\b_1}
+ (41)^{\a_4}{}_{\b_1} ~
\delta^{\a_1}_{\b_3} \delta^{\a_3}_{\b_2}
\delta^{\a_2}_{\b_4}
\non & &
+ (23)^{\a_2}{}_{\b_3} ~
\delta^{\a_4}_{\b_2} \delta^{\a_1}_{\b_4}
\delta^{\a_3}_{\b_1}
+ (34)^{\a_3}{}_{\b_4} ~
\delta^{\a_4}_{\b_2} \delta^{\a_1}_{\b_3}
\delta^{\a_2}_{\b_1} \Bigr) 
 \ .
\label{ fourpointcolord}
\fqr 
This vertex is represented in figure 4. 

\begin{figure}
\begin{center}
\epsfig{file=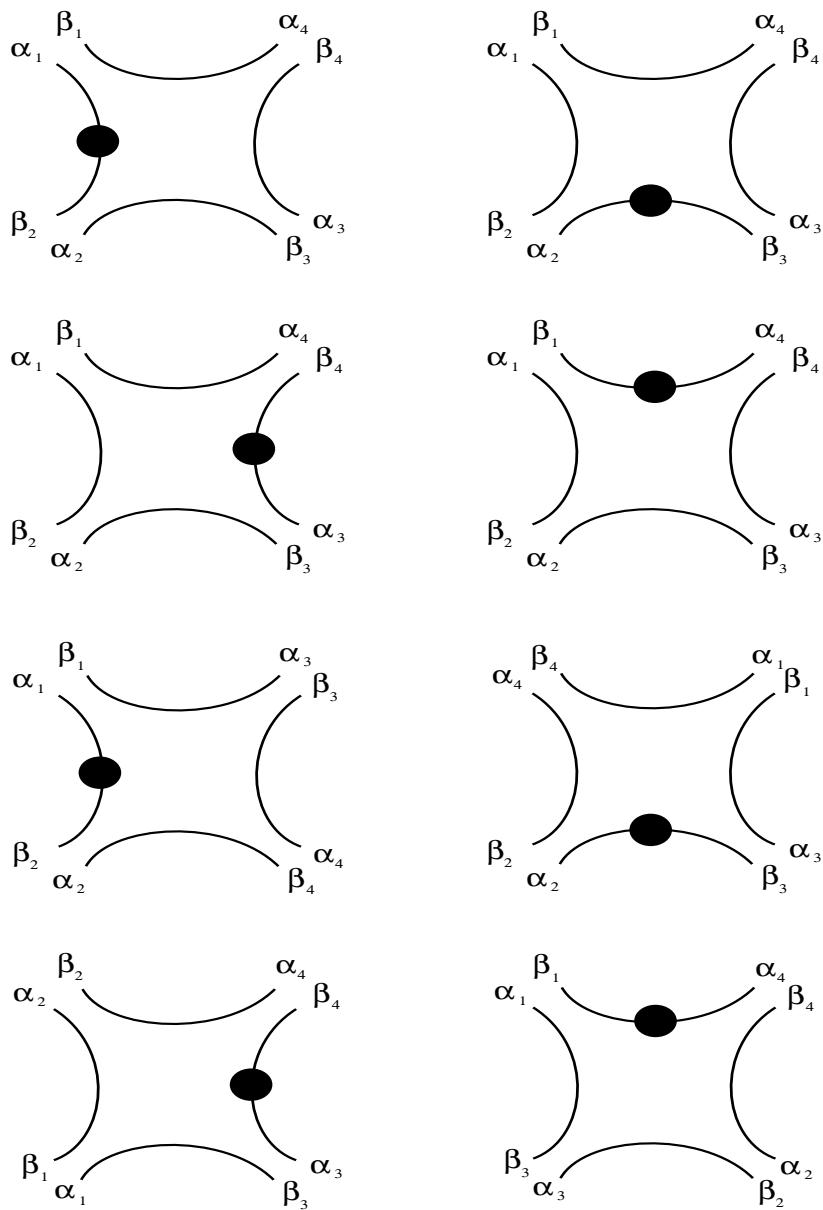,height=16cm,width=11cm}
\end{center}
\caption{Spin ordered diagrams of the color-ordered four-point vertex.}
\end{figure}

We end our discussion of the Feynman rules with the couplings of the gauge
bosons with the fermions; the simplest example is the coupling between two
$G$-fields and two fermions.  From the four-point interaction,
\be
{\cal L}_{GG\psi\psi} = {1\over 2M^2}
 \bigl( \partial_\gamma{}^{\da} G_a^{\a\gamma} \bigr)
 \bigl( \partial_{\rho\da} G_{b,\a}^{\rho} \bigr)
~ \psi^\delta \left\{ T_a^\psi, T_b^\psi\right\} \psi_\delta   \ ,
\ee
we obtain the Feynman rule
\be
\langle G_a^{\a_1}{}_{\b_1}(k_1) G_b^{\a_2}{}_{\b_2}(k_2)
 \psi^{\a_3}_i(k_3) \psi^{\a_4}_j (k_4) \rangle =
{1\over 2M^2} ~
k_1^{\a_1\da} k_{2,\b_1\da} ~\delta^{\a_2}_{\b_2}
C^{\a_3\a_4} ~\left\{ T_a^\psi, T_b^\psi\right\}_{ij} \ .
\ee
Another example vertex at the five-point level is obtained from the
term
\be
{\cal L}_{G^3\psi^2} =
{1\over M^4} \epsilon^{abd} ~
 \bigl(\partial_\gamma{}^{\da} G_a^{\a\gamma}\bigr)
G_b^{\a\b}
 \bigl( \partial_{\rho\da}  G_c^{\b\rho} \bigr)
 \psi^\delta \left\{ T_d^\psi, T_c^\psi\right\} \psi_\delta
\ee
There are many other vertices describing the couplings of fermions and
$G^{\a\b}$ fields found from the
third order  Lagrangian in eq.~(\ref{thirdord}) which are easily obtained, and
we do not list.  A straightforward expansion of the Lagrangian in eq.\ 
(\ref{nondualform}) generates these terms through the 
expansion of the inverse.  An important feature of all of the vertices to any
order is that they contain two powers of momentum contracted
in the fashion $k_i^{\a\da} k_j{}^\gamma{}_{\da}$.  

\section{Four-Point Massive Vector Amplitude}

In this section we give a derivation of a four-point tree amplitude 
between massive vector bosons.  We first introduce the spin trace, 
defined by
\be
{\rm Tr}[AB\cdots C]= A_{\a}{}^\b B_\b{}^\gamma \ldots
C_\rho{}^\a \ .
\ee

There are three topologies of Feynman diagrams which contribute to the
four-point tree amplitude.  The external line 
factors for the massive fields, $L_i^{\a\b}$, are described in 
section 2 and have a simple form in the rest frame.  These matrix line 
factors must be traced over in the calculations relevant to the four-point 
massive vector scattering.  The first set of Feynman diagrams contains 
naively a pole in the $s_{12}$ channel; the diagrams give the result
\beqs 
A_I(k_1,k_2,k_3,k_4) &=& {1\over M^2}{1\over s_{12}+M^2}~ 
\non & &  
\times {\rm Tr} \Bigl( \left[
 L_1  (12) L_2 - (1+2,1) L_1 L_2 - L_1 L_2 (2,1+2)
 \right]
\non & &  
\times \left[ L_3 (34) L_4 - L_3 L_4 (4,3+4) - (3+4,3) L_3 L_4
  \right] \Bigr)  \ .
\eeqs
Its counterpart with a cupped propagator leads to a multiple trace
structure,
\beqs 
A_I^{t} (k_1,k_2,k_3,k_4) &=& {1\over M^2} {1\over s_{12}+M^2}~ 
\non & &  
\times  {\rm Tr} \left(
 L_1  \left(12\right) L_2 - \left(1+2,1\right) L_1 L_2 - L_1 L_2
\left(2,1+2\right)
\right)
\non & &  
\times {\rm Tr} \left( L_3 \left(34\right) L_4 - L_3 L_4 \left(4,3+4\right)
 - \left(3+4,3\right)  L_3 L_4
  \right)  \ .
\eeqs 
The second class of diagrams contains a pole in the $s_{23}$-channel;
their contribution is
\beqs 
A_{II}(k_1,k_2,k_3,k_4) &=& {1\over M^2}{1\over s_{23}+M^2}~ 
\non & &   
\times {\rm Tr} \Bigl( \left[
 L_1 (14) L_4 - (1+4,1) L_1 L_4 - L_1 L_4 (4,1+4)
  \right]
\non & &   
\times \left[
 L_3 (32) L_2 - L_3 L_2 (2,2+3) - (2+3,3) L_3 L_2
  \right]  \Bigr) \ .
\eeqs
Their twisted counterparts are
\beqs 
A_{II}^t (k_1,k_2,k_3,k_4) &=& {1\over M^2}{1\over s_{23}+M^2}~ 
\non & &  
\times {\rm Tr} \left[  L_1 (14) L_4 - (1+4,1) L_1 L_4 - L_1 L_4 (4,1+4)
  \right]
\non & &  
\times {\rm Tr} \left[
L_3 (32) L_2 - L_3 L_2 (2,2+3) - (2+3,3) L_3 L_2
  \right] \ .
\eeqs 
The remaining contribution to the four-point tree amplitude comes from the
four-point vertex, which contains twelve terms:
\beqs  
A_{III} & =&  {1\over M^2} {\rm Tr} \left[ L_1 (12) L_2 L_3 L_4 + L_1 L_2
L_3 L_4 (41)
 + L_1 L_2 L_3 (34) L_4 + L_1 L_2 (23) L_3 L_4 \right]
\non & &  \hskip -.5in 
+ {1\over M^2} {\rm Tr} \left[ L_1 (12) L_2 L_4 L_3 + L_1 L_3 L_2 L_4 (41)
 + L_2 (23) L_3 L_1 L_4 + L_1 L_3 (34) L_4 L_2 \right] \ .
\eeqs 
The sum of these three combinations generates the spin-ordered (and 
color-ordered) massive four-point vector amplitude.  The line factors 
associated with the external states in \rf{dualpol} are required for the 
complete covariant result.  

We complete this section with the calculation of the massive analog of 
the maximally helicity violating process, i.e. scattering of the massive 
vector bosons possessing the same helicity $G^\pm$.  We write the momenta 
of the external vectors in terms of null momenta as $p_j=k_j+q$, with $q$ 
the same for all of the distinct bosons.  The line factors for the out-going 
$G^-$ states are 
\bqr  
L_{j,\a\b}^- = i {q_\a q_\b\over \langle qj\rangle} \ .  
\fqr 
Inspecting the contributions in equations (6.2) to (6.6) we see that all of 
the terms contain a contraction of $q^\a q_\a$ and are individually zero.  
The tree-level helicity process for this configuration of the massive 
gauge bosons is equal to zero (and similarly at $n$-point).  

\section{Vectorial Spin Ordering}

In the vector theory in the previous section we examined the
Feynman rules and amplitudes within the double line representation
of the spin algebra.  Because $G^{\a\b}$ has three components 
and transforms like a $3$-vector, it is suitable to describe the 
interactions in vector notation, in which the Feynman rules are made 
simpler.  We shall reformulate in this section 
the interactions in this notation.  

\begin{figure}[ht]
\begin{center}
\epsfig{file=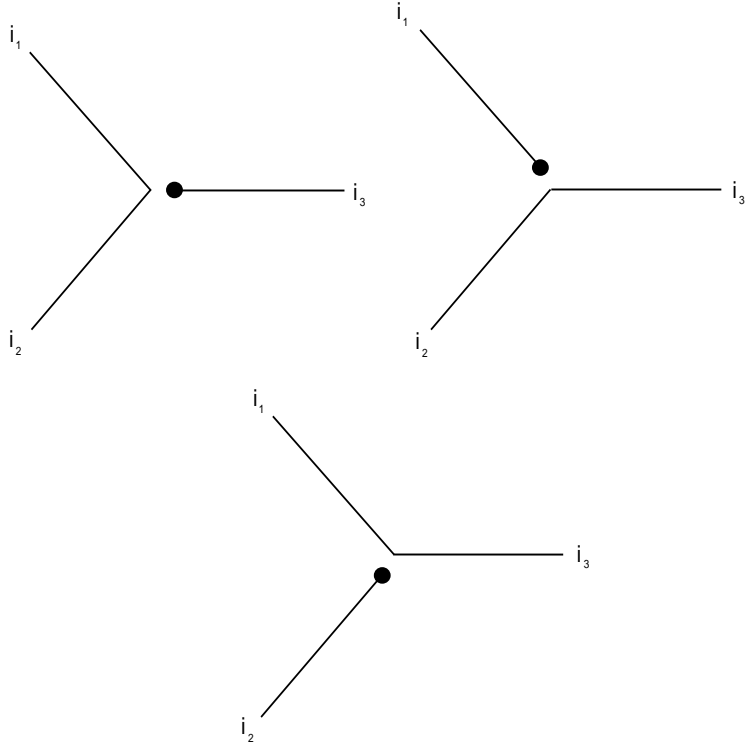,height=10cm,width=10cm}
\end{center}
\caption{The three terms of the three-point vertex in the single line 
representation.}
\end{figure}

We start by noting that a symmetric matrix $M_{\a\b}$ 
has three components and may be written as a three-vector, $M^i$.
We need to reexpress the trace formula in the previous section into
three-vector notation.  Using the identity
\be 
2 A_\a{}^\b B_{\b\gamma} = A_{(\a\vert}{}^\b 
 B_{\b\vert\gamma)}
+A_{[\a\vert}{}^\b B_{\b\vert\gamma]} \ ,
\label{symmanti}
\ee
we find a version of the trace formula
\be 
{\rm Tr}(ABCD)= A\cdot B C\cdot D - A\cdot C B\cdot D + A\cdot D B\cdot C 
\ , 
\ee
where we may translate into three vector notation with $A\cdot B=A^i B^i$
and
\be 
A^{\a\b}B_{\a\b} \rightarrow {1\over 2}A^i B_i \ . 
\ee
The Feynman rules in three-vector notation are obtained by translating
the previously derived ones.

The three-point color-ordered vertex in three-vector notation translates
into, after contracting with the polarization vectors $L_{\pm,0}$,
\be 
V_3= - {1\over M}\left[ L_1\cdot L_2 L_3 \cdot \left(12\right)
+ L_1\cdot L_3 L_2 \cdot \left(13\right)
+ L_2\cdot L_3 L_1 \cdot \left(23\right) \right] 
\ee
or in uncontracted form,
\be 
V_3^{i_1i_2i_3} = -{1\over M}\left[ \delta^{i_1i_2} \left(12\right)^{i_3}
 +  \delta^{i_1i_3} \left(13\right)^{i_2}
+ \delta^{i_2 i_3} \left(23\right)^{i_1} \right] \ .  
\label{vecthreept} 
\ee
where we have suppressed the helicity dependence in the line factors
in eq.\ (\ref{vecthreept}).  This vertex has a particularly simple form
when compared with the usual formulations.  The vertex is illustrated
graphically using the single line representation in figure 5; the dot
denotes the contraction $(ij)$ of momenta from the two adjacent lines.

We may find a similar form for the color-ordered four-point
vertex; this vertex becomes in three-vector notation, after contracting with
the external polarizaton vectors,
\bqr  
V_4 &=& -{2\over M^2}  \left[ L_1\cdot L_2~ (L_3\times L_4)\cdot(34)
 + L_2\cdot L_3~ (L_4\times L_1)\cdot(41) \right]
\non & &
 -{2\over M^2} \left[ L_3\cdot L_4 ~(L_1\times L_2)\cdot(12)
 + L_4\cdot L_1 ~(L_2\times L_3)\cdot(23)  \right]
\fqr    

The four-point vertex has only four terms and is a reduced
version of that appearing in the double-line representation.
In an uncontracted form, we have the color-ordered four-point
vertex written as
\bqr  
V_4^{i_1i_2i_3i_4} &=& -{2\over M^2} \left[ \delta^{i_1i_2} 
 ~ \epsilon^{i_3 i_4 i} (34)_i   
 + \delta^{i_2i_3} ~\epsilon^{i_4 i_1 i} (41)_i \right]
\non &&   
-{2\over M^2} \left[ \delta^{i_3i_4}~ \epsilon^{i_1 i_1 i} (12)_i
+ \delta^{i_4i_1} ~\epsilon^{i_2 i_3 i} (23)_i \right] 
\fqr  
This vertex is illustrated in the single line representation in figure 6.

\begin{figure}
\begin{center}
\epsfig{file=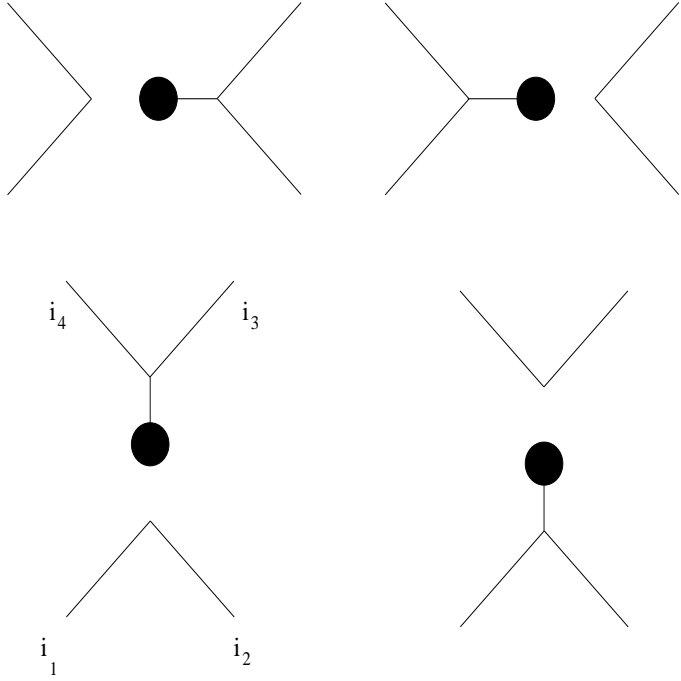,height=9cm,width=9cm}
\end{center}
\caption{The four terms contributing to the color-ordered four-point vertex 
in the single line representation.  The dot denotes contraction with the 
product $\epsilon^{abi}(mn)_i$ of the opposing two lines $m$ and $n$.}
\end{figure}

The forms of these vector indexed vertices is particularly simple
and imply a similar simplification of the three- and four-point
double line graphs; their form after converting the vector
line into a bi-spinor one gives the three and four-point vertices
illustrated in figures 7 and 8.  We have essentially enlarged
the vector into a bispinor label; the contraction of the
momenta is denoted by the dot on the line.  These forms are
equivalent to the  previous double line representations.

\begin{figure}
\begin{center}
\epsfig{file=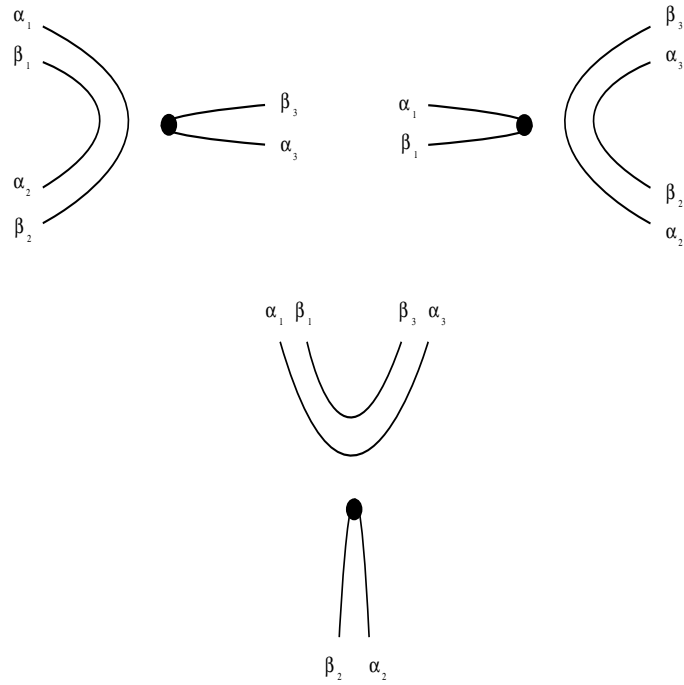,height=9cm,width=9cm}
\end{center}
\caption{The alternative double line representation of the three-point 
vertex.}
\end{figure}

\begin{figure}
\begin{center}
\epsfig{file=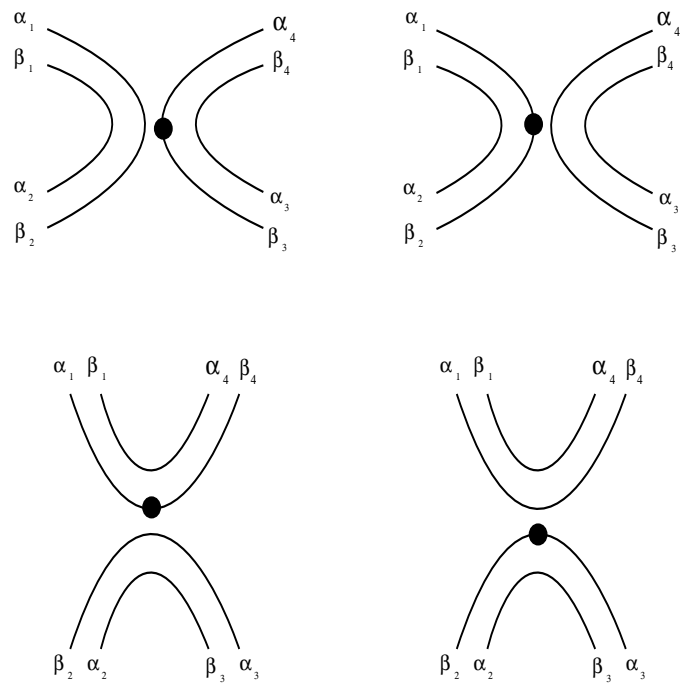,height=9cm,width=9cm}
\end{center}
\caption{The alternative double line representation of the 
four-point vertex.}
\end{figure}
 
\section{Cross-Sections} 

We conclude this section with an explanation on how to obtain the 
cross sections from
the amplitudes derived using the dualized Feynman rules.  First,
we denote the three independent polarizations in a three-component
bi-spinor form, $L_{\a\b}^\lambda\rightarrow
L_{\a\b}^{\gamma\delta}$.  Using this notation, our
helicity amplitudes are written as
\be 
A(k_1,\lambda_1;k_2,\lambda_2,\ldots)\rightarrow
A^{\a_1\b_1,\a_2\b_2,\ldots}(k_1,k_2,\ldots) \ .
\ee
There are two ways to make the complex conjugate and the 
corresponding cross-sections.  First, we may use the actual 
complex conjugation of our amplitude in the form,
\be 
A^\ast(k_1,\lambda_1;k_2,\lambda_2,\ldots)\rightarrow
A^{\da_1\db_1,\da_2\db_2,\ldots}(k_1,k_2,\ldots)
\ ,
\ee
The amplitude squared and summed over the first $p$ polarizations
must be contracted with a vector containing both undotted and
dotted indices.  The only vectors available are the momenta
$k_{i,\a\da}$.  The squared amplitude which is summed over the
first $m$ polarizations is contracted in the Lorentz covariant manner,
\bqr 
\sum_{\rm pol} \vert A\vert^2 &=&
 A^{\a_1\b_1 \ldots \a_n\b_n}(k_i) A^{\da_1\db_1
\ldots \da_n\db_n}(k_i) 
\non & & \times \left[ 
\left( {k_{1,(\da_1\vert(\a_1}\over M} {k_{1,\b_1)\vert\db_1)}\over M} \right) 
  \cdots \left( 
{k_{m,(\da_m\vert(\a_m}\over M} {k_{m,\b_m)\vert\db_m)}\over M} \right) 
 \right] \ , 
\fqr 
following from the orthogonality over the intermediate spin states 
$\sum_{\lambda=\pm,0} L_{\lambda}^{\a\b} L_{-\lambda}^{\da\db}$.  

In the completely chiral form the dualized Lagrangian defines first 
our amplitude $A^{\a_1\b_1,\a_2\b_2\cdots}$; the complex 
conjugate amplitude is found from the action written in terms of 
$G^{\da\db}$ together with flipping the helicity states 
from the external line factors.  This results in the same form as previously 
(but with all $i$'s exchanged with $-i$'s).  We denote this amplitude 
${\tilde A}^{\a_1\b_1,\a_2\b_2\cdots} (k_1,\lambda_1;k_2, 
\lambda_2,\ldots)$.  Then the cross-section is derived by the orthogonality 
relation in \rf{orthothree} to join $A$ together with ${\tilde A}$: 
\be  
\sum_{\rm pol} \vert A\vert^2 =
 A^{\a_1\b_1 \ldots \a_n\b_n}(k_i) {\tilde A}^{ 
 {\tilde\a}_1{\tilde\b}_1
\ldots {\tilde\a}_n{\tilde\b}_n}(k_i) \left[ 
 \left( C_{({\tilde\a}_1\vert(\a_1} C_{\b_1)\vert{\tilde\b}_1)} \right) 
\cdots 
 \left( C_{({\tilde\a}_m\vert(\a_m} C_{\b_m)\vert{\tilde\b}_m)} 
\right)
\right] \ , 
\ee 
where the tilde denotes a separate index labelling the same 
representation as the untilded indices.  The latter definition of 
the cross-section is simpler to implement in practice as only one 
type of index is used and spin ordering is naturally useful.

All final states are summed over in this manner.  We may also perform 
the average over initial states by contracting and summing over their 
polarizations.

\section{Electroweak Model}

\subsection{Dualization}

In this section we will consider the well known $SU(2)_L\times U(1)$
electroweak portion of the Standard Model.  The dualized Lagrangian is
derived after integrating out as in the previous section the gauge fields
for the massive vector bosons from a first order from of gauge theory.

We first review the construction of this model. The Lagrangian
for the $SU(2)_L\times U(1)$ theory contains the gauge and
scalar sectors,
\be 
{\cal L}_g =
  {1\over 2} F^{\a\b} F_{\a\b}
 + {1\over 2} F^{a,\a\b} F_{a,\a\b}
\ ,
\label{orig}
\ee
and
\be 
{\cal L}_s= \nabla_{\a\da} \Phi^\dagger
\nabla^{\a\da}\Phi - \mu^2 \Phi^\dagger \Phi
+~\lambda \left[\Phi^\dagger\Phi\right]^2
\ .
\ee
The scalar covariant derivative weighted with the
appropriate hypercharge assignments is
\be 
\nabla_{\a\da}=\partial_{\a\da} +
i g_2 A_{\a\da}^a \left( {t_a\over 2}\right)
-{i\over 2} g_1 A_{\a\da} \ ,
\ee
where the scalar field is in doublet form,
\be 
\Phi= \pmatrix{ \phi_1 \cr \phi_2}  \ .
\ee
Furthermore, we may add doublet fermions through their minimal 
couplings to the gauge fields, ${\cal L}_f= -{\bar\Psi}^{\da} 
i\nabla_{\a\da} \Psi^\a \ .$  The generation of an 
electron and its neutrino, for example, is
\be 
\Psi= \pmatrix{ \psi_e \cr \psi_{\nu_e} } \ .
\ee
We may couple further generations to the gauge fields.

The standard form of the electroweak model is obtained by going first
to a unitary gauge and then making a field redefinition which
expresses the massive vector fields $W^{\pm}$ and $Z$ in terms of the
massive gauge fields $A^a$ and $B$. In obtaining the dualized  theory
there is freedom in making the various steps to obtain the new
Lagrangian.  Furthermore, we shall use a Stueckelberg mass term for
the photon to treat this field on the same footing as the $W$ and $Z$
fields; the amplitudes obtained for the massless photon case may be
found by taking the massless limit.  (The scalar component of the
massive vector must  decouple from the S-matrix in this limit.)  The
procedure we adopt in obtaining our formulation of the electroweak
model is described by: (1) writing  a first order form to
eq.~(\ref{orig}), (2) choose a unitary  gauge, (3) integrate out the
massive vector bosonic gauge fields.  One may interchange steps (1)
and (2) in deriving the final form for the Lagrangian.

The first order form to the electroweak model is found by introducing
into the Lagragian in  eq~(\ref{orig}) the field strengths
$G_{\a\b}^A$ and
$G_{\a\b}^B$,
\be 
 {1\over 2} F^{a,\a\b}  F_{a,\a\b}
 ~\Rightarrow~
 -{1\over 2} G^{a,\a\b}G_{a\a\b}
 + G^{a,\a\b}F_{a,\a\b} \ ,
\ee 
and
\be 
 {1\over 2} F^{\a\b}  F_{\a\b}
  ~\Rightarrow~
 -{1\over 2} G^{\a\b}G_{\a\b}
+ G^{\a\b}F_{\a\b} \ .
\ee
The unitary gauge is specified
by using the $SU(2)_L$ gauge freedom to choose the gauge $\Phi=
{1\over\sqrt{2}} (\phi+v) I$.    We further use the standard basis for the
gauge
fields $W^\pm_{\a\da}$, $Z_{\a\da}$, and photon
$A_{\a\da}$ fields:
\be 
Z_{\a\da} = {g_2 A_{\a\da}^3 - g_1 A_{\a\da}
\over [g_1^2+g_2^2]^{1/2}}, \qquad
B_{\a\da} = {g_1 A_{\a\da}^3 + g_2 A_{\a\da}
\over [g_1^2+g_2^2]^{1/2}}, \qquad
W_{\a\da}= {1\over\sqrt{2}}\left[A_{\a\da}^1
\pm i A_{\a\da}^2\right]  \ .
\label{standbasis}
\ee
In the unitary gauge the scalar kinetic term becomes,
\be 
\nabla_{\a\da} \Phi^\dagger \nabla^{\a\da}\Phi =
 {1\over 2}  \partial_{\a\da} \phi \partial^{\a\da} \phi
\ee 
\be 
 + {g_2^2\over 4} \left(v+\phi\right)^2 W^+\cdot W^-
+ {1\over 8} (g_1^2+g_2^2) \left(v+\phi\right)^2  Z\cdot Z
\ .
\ee
The pure Higgs couplings give through the $\Phi^4$ potential we have chosen,
\be 
{\rm Tr}~ {\lambda\over 4} (\Phi^2-v^2)^2 =  {1\over 2} m_h^2 \phi^2
+{\lambda\over 4} \left( 4v\phi^3+\phi^4\right)  \ .
\ee
Next we must integrate out the combinations of fields $A_{\a\da}^a$
and $B_{\a\da}$ that correspond to the $W^\pm$ and $Z$ field.  We
shall reexpress the first-order form of the Lagrangian in these gauge field
variables.

First we examine the contributions from the dualized field strengths.
In terms of the redefined fields we have from the $U(1)$-field contribution,
\be 
G^{\a\b}F_{\a\b} = i\kappa~ G^{\a\b}
 \partial_{\a\dt\gamma}
\left(g_2 B-g_1 Z\right)^{\dt\gamma}{}_\b
\ee
where $\kappa=1/ \sqrt{g_1^2+g_2^2}$.  The remaining contributions
found from inverting the basis in eq.~(\ref{standbasis}) are expressed as
\beqs 
G^{a,\a\b}   F^a_{\a\b} &=&
 G^{1,\a\b} \Bigl\{
 {i\over\sqrt{2}} \partial_{\a\dt\gamma} (W^+ +W^-)^{\dt\gamma}{}_\b
 +{`\kappa\over\sqrt{2}} ~(g_2 Z+g_1 B)_{\a\dt\gamma}
  (W^+ - W^-)^{\dt\gamma}{}_\b \Bigr\}
\non & &
+ G^{2,\a\b} \Bigl\{
 {1\over\sqrt{2}} \partial_{\a\dt\gamma} (W^+ -W^-)^{\dt\gamma}{}_\b
 -{i\kappa\over\sqrt{2}}~ (g_2 Z+g_1 B)_{\a\dt\gamma}
  (W^+ + W^-)^{\dt\gamma}{}_\b \Bigr\}
\non & &  
+ G^{3,\a\b} \Bigl\{
 {\kappa} ~\partial_{\a\dt\gamma}
 (g_2 Z+g_1 B)^{\dt\gamma}{}_\b
 -{1\over 2} (W^+ + W^-)_{\a\dt\gamma}
  (W^+ - W^-)^{\dt\gamma}{}_\b \Bigr\} \ .
\nonumber \\ 
\label{basechange}
\eeqs 
We shall also use a new basis for the $G$-fields,
$$ 
G_B^{\a\b}={1\over \sqrt{g_1^2+g_2^2}} (g_1 G_3^{\a\b}
 + g_2 G^{\a\b}), \qquad
G_Z^{\a\b}= {1\over
\sqrt{g_1^2+g_2^2}} (g_2 G_3^{\a\b}-g_1 G^{\a\b}),
$$ 
\be 
G_{W^\pm}^{\a\b} = {1\over \sqrt{2}}( G_1^{\a\b} \pm
i G_2^{\a\b}) \ .
\ee 
Before integrating out the massive fields we write the above in the
compact notation which is useful for the Gaussian functional integration;
furthermore, we add the mass term for the photon field,
$$  
G^{a,\a\b} F^a_{\a\b} + G^{\a\b}F_{\a\b}
+ {g^2\over 4} \left( v+\phi\right)^2  W^+\cdot W^-
+ {1\over 8} (g_1^2+g_2^2) \left(v+\phi\right)^2  Z\cdot Z +
{m^2\over 2} B^{\a\da} B_{\a\da}
$$
\be  
 = \left(\partial_{\a\dt\gamma} {\tilde G}^{i,\a}{}_\b \right)
  ~Z^{j,\b\dt\gamma}
 + Z^{i~\dt\gamma}_\a ~\Bigl( G_k^{\a\b} M_{ij}^k
 + Q_{ij} C^{\a\b} \Bigr)~
 Z^j_{\b\dt\gamma}
\label{matreq}
\ee 
where $Z= (W^+,W^-,Z,B)$ and ${\tilde G}=(G_{W^+}, G_{W^-},
G_Z,G_B)$.
The matrices $M$, $N$,  and the vector $P$ are  determined from the
expansion in eq.~(\ref{basechange}).

The functional integration over the set of $Z_j$ fields in eq.~(\ref{matreq})
is Gaussian and may be performed in a straightforward fashion.  We obtain
\be 
{\cal L}= V^{\a\dt\gamma}_i~ \Bigl[{1\over X}\Bigr]_{\a\b}^{ij}
~ V^{\b}_{j}{}_{\dt\gamma}
- {1\over 2} G_Z^{\a\b} G_{Z,\a\b} - {1\over 2} G_B^{\a\b}
G_{B\a\b} - G_{W^+}^{\a\b} G_{W^-,\a\b}
\ee 
where
\be 
V_i^{\a\dt\gamma}= \partial_{\gamma}{}^{\dt\gamma} {\tilde
G}^{\gamma\a}_i
\ee
and
\be 
X^{ij}_{\a\b} = G_{k,\a\b} M_{ij}^k
  + Q_{ij} C_{\a\b} \ .
\ee
The matrix $G_k^{\a\b}M^k_{ij}$ is
$$
G_k^{\a\b} M_{ij}^k =
$$
\beqs 
 \pmatrix{ 0 &  {\kappa\over 2}(g_2 G_Z+g_1G_B)^{\a\b} &
 -{g_2\kappa\over 2} G_{W^-}^{\a\b}
 & -{1\over 2} g_1\kappa G_{W^-}^{\a\b}
\cr
-{\kappa\over 2}(g_2 G_Z+g_1G_B)^{\a\b}  & 0
 & {g_2\kappa\over 2} G_{W^+}^{\a\b}
 & {1\over 2} g_1\kappa G_{W^+}^{\a\b}
\cr
 {g_2\kappa\over 2} G_{W^-}^{\a\b}
 & -{g_2\kappa\over 2} G_{W^+}^{\a\b} & 0 & 0
\cr
 {1\over 2} g_1\kappa G_{W^-}^{\a\b} &
 -{1\over 2} g_1\kappa G_{W^+}^{\a\b} & 0 & 0
}  \ .
\label{gmmatrix}
\eeqs 
The mass matrix $Q_{ij}$ is similarly obtained from the spontaneously
broken contribution in eq.~(\ref{matreq}) and is
\be 
Q_{ij} = {1\over 2}
 \pmatrix{ 0 & {1\over 4}\left( v+\phi\right)^2  {g^2_2} & 0 & 0
\cr {1\over 4}\left( v+\phi\right)^2 {g^2_2}  & 0 & 0 & 0
\cr 0 & 0 &  {1\over 4}\left( v+\phi\right)^2 (g_1^2+g_2^2) & 0
\cr 0 & 0 & 0 & m^2 } \ ,
\label{qmatrix}
\ee
where the Stueckelberg mass parameter is $m^2$.
The Feynman rules are obtained after perturbatively inverting the matrix
$X$ around the mass parameters of the $G_{W^\pm}$, $G_Z$, and $G_B$ fields.
Note that we must add in a mass term for the photon at an intermediate 
stage to perform the dualization; we take $m=0$ at the 
end of the S-matrix calculations.  

\subsection{Feynman Rules}

In this section we give the Feynman rules for the dualized
electroweak model presented in the previous secton.  The matrix
$X_{ab}^{\a\b}$ has the same form as in eq.~(\ref{Xmatrixone}); the
perturbative inverse to first order is
\beqs 
\Bigl[ {1\over X} \Bigr]_{\a\b}^{ij} &=& 
\bigl[ C_{\a}{}^{\gamma} Q^{-1}_{ik} \bigr]
 \bigl[ C_{\gamma\b} \delta_{kj} -
 (CQ)^{-1}_{kl}{}_\gamma{}^{\rho}~ G_{(a)\rho\b} M^{(a)}_{lj} + \ldots
\bigr]
\\ & &
 = C_{\a\b} Q^{-1}_{ij} +
 Q^{-1}_{ik} Q^{-1}_{kl} G_{\a\b(a)} M^{(a)}_{lj}
+ \ldots \ .
\label{firstorderelec}
\eeqs 
where the inverse to the matrix $Q_{ij}$ is
\beqs 
Q^{-1}_{ij} = 2
 \pmatrix{ 0 & {4\over (v+\phi)^2} {1\over g_2^2} & 0 & 0
\cr {4\over (v+\phi)^2}{1\over g_2^2} & 0 & 0 & 0
\cr 0 & 0 & {4\over (v+\phi)^2}{1\over (g_1^2+g_2^2)} & 0
\cr 0 & 0 & 0 & {1\over m^2} }  \ .
\label{inverseQ}
\eeqs 
This expansion generates couplings to third order in the massive
${\tilde G}_j^{\a\b}$ fields.  We may also expand in the scalar
field $\phi$
around the vacuum value $v$ in eq.~(\ref{inverseQ}).

Explicitly the $V_i^{\a\da}$ vector is
\beqs 
V^{\a\dt\gamma}_j = \pmatrix{
\partial_\gamma{}^{\dt\gamma} G_{W^+}^{\gamma\a}
\cr \cr
\partial_\gamma{}^{\dt\gamma} G_{W^-}^{\gamma\a}
\cr \cr
\partial_\gamma{}^{\dt\gamma} G_Z^{\gamma\a}
\cr \cr
\partial_\gamma{}^{\dt\gamma} G_B^{\gamma\a}}
\eeqs
The lowest order term in eq.~(\ref{firstorderelec}) generates the
propagators,
\bqr  
{\cal L}_0^p &=& {8\over v^2(g_1^2+g_2^2)}
~G_{Z,\a\b}\left(\Box-m_Z^2\right) G_Z^{\a\b}
 + {16\over v^2g_2^2}~ G_{W^+,\a\b}\left(\Box-m_W^2\right) G_{W^-}^{
\a\b}
\non & &  
+ {2\over m^2} G_B^{\a\b} \left( \Box - m_B^2 \right) G_{B,\a\b}
\ ,
\fqr  
where the masses are
\be 
m_W^2= {v^2g_2^2\over 16} \qquad m_Z^2= {v^2(g_1^2+g_2^2)\over 16}
\qquad m_B^2 = {1\over 4} m^2 \ .
\ee 
We have used the fact that
$\partial_\a{}^{\dt\gamma}\partial_{\b\dt\gamma}
= -C_{\a\b} \Box$.

The interactions to third order in ${\tilde G}_j^{\a\b}$ require the
first
correction in eq.~(\ref{firstorderelec}).  We need the following
matrix form of the first correction,
$$ 
Q^{-1}_{ik} Q^{-1}_{kl} G_{\a\b(a)} M^{(a)}_{lj} = 
$$ 
\beqs 
\pmatrix{ 0 & \a (g_2 G_Z+g_1G_B)^{\a\b}
 & \b_1 G_{W^-}^{\a\b} & \gamma_1 G_{W^-}^{\a\b}
\cr
-\a (g_2 G_Z+g_1G_B)^{\a\b}  & 0
 & -\b_1 G_{W^+}^{\a\b} & -\gamma_1 G_{W^-}^{\a\b}
\cr
\b_2 G_{W^-}^{\a\b} & -\b_2 G_{W^+}^{\a\b} & 0 & 0
\cr \gamma_2 G_{W^-}^{\a\b} & -\gamma_2 G_{W^-}^{\a\b} & 0 & 0 }
\ ,
\eeqs
where
\bqr  
\a &=& {\kappa\over g_2^4}{32\over (v+\phi)^2} \qquad
\b_1 = -{\kappa\over g_2^3} {32\over (v+\phi)^2} \qquad
\b_2 = g_2 \kappa^5 {32\over (v+\phi)^2} \ ,
\non & & \hskip .5in 
\gamma_1 = -{1\over 2m^4} g_1\kappa \qquad
 \gamma_2 = {g_1\kappa\over g_2^4} {32\over (v+\phi)^2} \ .
\fqr  
Since the matrix $Q^{-1}_{ik} Q^{-1}_{kl}$ has only diagonal entries,
the matrix $Q^{-1}_{ik} Q^{-1}_{kl} G_{\a\b(a)} M^{(a)}_{lj}$ is
very simple.  There are several interactions at this order.  We
have the interactions involving only the $G$-fields,
\bqr  
{\cal L}_{G^3} & =& \a_1 \left( g_2G_Z+g_1G_B\right)_{\a\b} \left[
 \left( \partial_{\rho\dt\gamma} G_{W^-}^{\rho\a}\right)
 \left( \partial_\gamma{}^{\dt\gamma} G_{W^+}^{\gamma\b}\right)
 -  \left( \partial_{\rho\dt\gamma} G_{W^+}^{\rho\a}\right)
 \left( \partial_\gamma{}^{\dt\gamma} G_{W^-}^{\gamma\b}\right)
\right]
\non & &
+ (\b_1-\b_2) \left( \partial_\gamma{}^{\dt\gamma}
G_Z^{\gamma\b} \right)
\left[ \left( \partial_{\rho\dt\gamma} G_{W^+}^{\rho\a}\right)
G_{W^-,\a\b} -
 \left( \partial_{\rho\dt\gamma} G_{W^-}^{\rho\a}\right)
G_{W^+,\a\b} \right]
\non & &
+ (\gamma_1-\gamma_2) 
 \left( \partial_\gamma{}^{\dt\gamma} G_B^{\gamma\b} \right)
\left[ \left( \partial_{\rho\dt\gamma} G_{W^+}^{\rho\a}\right)
G_{W^-,\a\b} -
 \left( \partial_{\rho\dt\gamma} G_{W^-}^{\rho\a}\right)
G_{W^+,\a\b} \right]
 ,
\fqr  
We end this section with a brief discussion of the next order
terms.  The expansion of the inverse matrix in eq.~(\ref{firstorderelec}) to
second order yields generically the terms
\be 
{\cal L}_{G^4}= A^{ijkl} 
 \left(\partial_\gamma{}^{\dt\gamma} G_i^{\gamma\a}\right)
 G_{j,\a\rho} G_k^\rho{}_\b 
\left(\partial_{\mu\dt\gamma} G_l^{\mu\b}\right) \ ,
\ee
where $G_i=\left( G_{W^+},G_{W^-},G_Z,G_B\right)$, and the coefficients 
of the interactions and the rules can be determined.  The couplings of 
the $SU(2)_L$ duals has the same form as the vertex in eq.~(\ref{couplings}).

\section{Conclusions} 

In this work we have explored the selfdual inspired reformulation 
of massive vector theories.  These manipulations have several advantages: 
We have a new spin-ordering analogous to color flow and the Feynman rules 
are suitable (when coupled to matter) to rederiving amplitudes closer to 
maximally helicity violating ones in an efficient manner.  

We have presented the selfdual reformulation for a spontaneously broken 
gauge theory (and Stuekelberg models as examples) with general matter 
content.  Furthermore, the example of the electroweak model is analyzed.  The 
Lagrangians contain an infinite number of terms: This arises from the 
reshuffling of the perturbative expansion at arbitrary order.  
Although the resulting action is more complicated, 
the Feynman diagrams are simpler,
since manipulations normally performed repeatedly at each propagator
and vertex in each diagram have been performed once and for all
in the corresponding terms in the action.
In the massless 
limit the expansions are formulated 
around the MHV (or selfdual) configurations and allows for a perturbative 
analysis in helicity flips at arbitrary $n$-point.  

It would be interesting to demonstrate the use of these rules in practical 
applications for massive vector processes at higher order.  This would entail 
a calculation (i.e.\ $6$-point) or further work at loop-level.  As the tensor 
algebra is reduced and spin-ordered (the diagrams written in smaller gauge 
invariant sets as well as labeled by only one chiral spinor index), the 
integral evaluations are the more complicated aspect.  In our approach
we have reduced the spin dependence of the vertices and interactions.
This aspect is the primary complication associated with a direct 
higher-than-four-point calculations.  

There are also interesting aspects to formulating spontaneously 
broken Yang-Mills theory with gauge group $H$ as a ${\rm dim} 
(\hbox{scalars})+3\times{\rm dim}(H)$ non-linear sigma model.  This 
dual formulation is chiral and written in terms of selfdual field 
strengths, an analog of holomorphy in $\cN=2$ super Yang-Mills theory.  
It might be interesting to interpret the isometries of this dual 
non-linear sigma theory within the conventional formulation of the 
gauge theory and also the topologically non-trivial gauge configurations 
and transformations.  

Selfdual gauge theory admits an infinite dimensional symmetry structure; 
as the former is a truncation of gauge theory \cite{Chalmers:1996rq} these 
conserved currents exist in a limit of the full Yang-Mills theory.  The selfdual 
approach may permit further progress in determining structure in the full gauge 
theory related to these conserved currents.  

\section*{Acknowledgements}
  
The work of G.C. is supported in part by the US Dept. of Energy, Division 
of High Energy Physics, Contract W-31-109-ENG-38.  The work of W.S. is 
supported in part by NSF Grant No. PHY 9722101. 

\appendix
\section{Standard Electroweak Model}

The bosonic sector of the electroweak model is listed below to facilitate 
the comparison with our work.  We break the bosonic contributions into
several terms, and in the following we denote $\una=\a\da$ 
and define the field $Y^{\a\da}$ by
\be 
Y_{\una} = \cos\theta_w Z_\una +\sin\theta_w B_\una \ .
\ee
The mixing angle is given by $\cos\theta_W=g_2/\left(g_1^2+g_2^2\right)^{1/2}$.

The propagators for the gauge fields are found from
\be 
{\cal L}_p = {1\over 2} F^{\a\b}F_{\a\b} + {1\over 2}
 F^{Z,\a\b}F^Z{}_{\a\b} + 
 F^{W^+,\a\b} F^{W^-}{}_{\a\b}
+{m_Z^2\over 2} Z\cdot Z
 + m_W^2 W^+\cdot W^-
\ee 
where the field strengths are defined in the manner $F^{W^+}_{\a\b} =
i\partial_{(\a}{}^{\dt\gamma} W^{+}{}_{\b)\dt\gamma}$; the mass
parameters
are
\be 
m_w={1\over 2}g_2v \qquad
m_z={1\over 2}(g_1^2+g_2^2)^{1/2} v \ .
\ee 
The tri-linear couplings of the gauge fields are found to be
\bqr  
{\cal L}_t &=& ig_2 \left( \partial_\una W^+{}_\unb - \partial_\unb
 W^+{}_\una\right)  W^{-,\unb} Y^\una
+ ig_2 \left( \partial_\una W^-{}_\unb - \partial_\unb
W^-{}_\una\right)
  W^{+,\unb} Y^\una
\non & &  
- ig_2 \left( W^-{}_\una W^+{}_\unb - W^-{}_\unb W^+{}_\una\right)
 \partial_\una Y_\unb
\ .
\fqr  
The quartic couplings are given by
\bqr  
{\cal L}_q &= & - g_2^2~ W^+\cdot W^-~ Y\cdot Y
+ g_2^2~ W^+\cdot Y ~ W^-\cdot Y
\non & &  
+{g_2^2\over 2}~ \left( W^+\cdot W^+ W^-\cdot W^- - 
  W^+\cdot W^- W^-\cdot W^+
 \right)
\ , 
\label{eleceqn}
\fqr  
There are also the contributions containing explicit couplings to the Higgs
particle.  With the $\phi^4$ potential we obtain, 
\bqr 
{\cal L}_\phi & =& {1\over 2}  \partial_\una \phi \partial^\una \phi -
{1\over 2}
m_\phi^2 \phi^2  -{\lambda\over 4} \left( 4v\phi^3+\phi^4\right)
\non & &  
 + {g^2\over 4} \left( 2v\phi+\phi^2\right)  W^+\cdot W^-
+ {1\over 8} (g_1^2+g_2^2) \left(2v\phi+\phi^2\right)  Z\cdot Z
\ . 
\fqr  
with a mass $m^2_\phi= 2\lambda v^2$.  The doublet fermions may
be added to the bosonic sector described above; we do not list
their couplings.

\end{document}